\documentclass[11pt]{article}
\usepackage{amsmath,amssymb}
\usepackage[T1]{fontenc}
\usepackage{lmodern}
\numberwithin{equation}{section}

\def\hi{\wh\imath}

\let\iy\infty
\let\op\oplus
\let\ot\otimes
\let\ov\overline

\let\pa\partial
\let\q\quad
\def\qh#1{\quad\hbox{#1}\quad}
\let\td\tilde
\let\tm\times
\let\ul\underline
\let\wh\widehat
\let\wt\widetilde
\let\a\alpha
\let\b\beta
\let\d\delta
\let\D\varDelta
\let\ve\varepsilon
\let\g\gamma
\let\G\varGamma
\let\ka\kappa
\let\la\lambda
\let\La\varLambda
\let\n\nabla
\let\o\omega
\let\O\varOmega
\let\F\varPhi
\let\vf\varphi
\let\Ps\varPsi
\let\PS\varPsi
\let\si\sigma

\let\t\theta
\let\T\varTheta
\let\z\zeta
\def\SU{\text{SU}}
\def\U{\text{U}}
\def\(#1){{(#1)}}
\def\[#1]{{[#1]}}
\def\Ft#1#2{\F^{#1}_{\td #2}}
\def\gv{\nad{gauge}}

\def\pl{_{\rm pl}}
\def\m{{\mu\nu}}
\def\Am{A_\mu}
\def\Bm{B_\mu}
\def\cD{\mathcal D}
\def\cL{\mathcal L}
\def\cLY{\mathcal L_{\rm YM}}
\def\cLf{\mathcal L_{\rm \fe s}}

\def\C{\mathbb C}
\def\R{\mathbb R}
\def\fg{\mathfrak g}
\def\fh{\mathfrak h}
\def\fm{\mathfrak m}
\def\tA{{\tilde A}}
\def\ta{{\tilde a}}
\def\tB{{\tilde B}}
\def\tb{{\tilde b}}
\def\tC{{\tilde C}}
\def\tc{{\tilde c}}
\def\dg #1,#2,#3,{{#1{}_{#2}}^{\!#3}}
\def\gd #1,#2,#3,{{#1{}^{#2}}_{\!#3}}
\def\gdg #1,#2,#3,#4,{{{#1{}^{#2}}_{\!#3}}{}^{\!#4}}
\def\dgd #1,#2,#3,#4,{{{#1{}_{#2}}^{\!#3}}{}_{\!#4}}
\def\nad#1#2{\overset{\rm #1}{#2}}

\def\falg{{\mathrel{\lower5pt\hbox{${\scriptstyle\sim}$}\hskip-5pt g}}}
\def\fal#1{{\mathrel{\lower4pt\hbox{${\scriptstyle\sim}$}\hskip-7.5pt #1}}}

\def\Dint{\overset{\rm (int)}{\hbox to6pt{\rm D\hss}\hskip-5pt\slash}}
\def\dwa#1{\textstyle{\substack{#1}}}
\newdimen\krop
\setbox0=\hbox{$\scriptstyle\a$}
\krop=\wd0
\def\cdt{\hbox to\krop{$\scriptstyle\hfil\cdot\hfil$}}
\def\lw#1 {\lower#1pt\hbox\bgroup$\scriptstyle}
\def\eg{$\egroup}
\def\hor{\operatorname{hor}}

\def\Ad{\operatorname{Ad}}
\def\Tr{\operatorname{Tr}}
\def\Spin{{\rm Spin}}
\def\SO{{\rm SO}}
\def\diag{{\rm diag}}
\def\pap#1#2{\frac{\pa #1}{\pa #2}}
\def\udots{\lower1pt\hbox to4pt{\hss$\cdot$\hss}\raise2pt\hbox to4pt{\hss$\cdot$\hss}\raise5pt\hbox to4pt{\hss$\cdot$\hss}}

\let\ti\textit
\def\beq#1 #2\e{\begin{equation}\label{#1}#2\end{equation}}
\def\bea#1 #2\e{\begin{align}\label{#1}#2\end{align}}
\def\bml#1 #2\e{\begin{multline}\label{#1}#2\end{multline}}
\def\bmlg#1\e{\begin{multline*}#1\end{multline*}}
\def\bg#1 #2\e{\begin{gather}\label{#1}#2\end{gather}}
\let\bal\aligned \let\eal\endaligned
\let\bga\gathered \let\ega\endgathered
\def\bma{\left(\begin{array}{c|c}} \def\ema{\end{array}\right)}
\def\bca{\begin{cases}}
\def\eca{\end{cases}}
\def\bit{\begin{itemize}}
\def\eit{\end{itemize}}
\def\ben{\begin{enumerate}}
\def\een{\end{enumerate}}
\let\nn\nonumber
\def\lb#1 {\label{#1}}
\let\er\eqref

\def\up#1{\uppercase{#1}}
\def\E{\expandafter\up}

\def\Ca{Clifford algebra}
\def\cf{coefficient}
\def\cm{composition}
\def\cfn{confinement}
\def\cn{connection}
\def\cv{conservation}
\def\ct{constant}
\def\cd{coordinate}
\def\co{cosmolog}
\def\ci{covariant}
\def\cvt{curvature}
\def\dv{derivative}
\def\di{di\-men\-sion}
\def\elm{electromagneti}
\def\el{element}
\def\e{equation}
\def\fe{fermion}
\def\fw{following}
\def\f{function}
\def\fn{fundamental}
\def\gn{generalization}
\def\gz{generalized}
\def\gr{gravitation}
\def\Hm{Higgs' mechanism}
\def\hc{hypercomplex}
\def\ia{interaction}
\def\iv{invarian}
\def\KK{Kaluza--Klein }
\def\KWK{Kerner--Wong--Kopczy\'nski }
\def\KC{Killing--Cartan }
\def\Lg{lagrangian}
\def\LC{Levi-Civita }
\def\Li{Lie algebra}
\def\MR{Moffat--Ricci}
\def\mo{morphi}
\def\nA{non-Abelian}
\def\JT{Nonsymmetric Jordan--Thiry Theory}
\def\nos{nonsymmetric}
\def\NK{Nonsymmetric Kaluza--Klein Theory}
\def\pc{particle}
\def\Pl{Planck}
\def\pt{potential}
\def\pp{principle}
\def\rp{represent}
\def\sf{satisf}
\def\so{solution}
\def\spt{space-time}
\def\spf{spinor field}
\def\sn{spontaneous}
\def\ssb{spontaneous symmetry breaking}
\def\sc{structur}
\def\st{such that }
\def\s{symmetr}
\def\tf{transformation}
\def\un{unification}
\def\v{variation}
\def\wrt{with respect to }
\def\YM{Yang--Mills}
\let\TM\texttrademark

\author{M. W. Kalinowski\\
Bioinformatics Laboratory, Medical Research Centre,
Polish Academy of Sciences,\\
02-106 Warszawa, Poland\\
e-mail: markwkal@bioexploratorium.pl, mkalinowski@imdik.pan.pl}
\title{Fermion Fields\\
in the (Non)Symmetric Kaluza--Klein Theory}
\begin{document}
\maketitle

\vskip20pt
\begin{abstract}
The paper is devoted to the unification of fermions within Nonsymmetric Kaluza--Klein Theories.
We obtain a Lagrangian for fermions in Non-Abelian Kaluza--Klein Theory and Non-Abelian
Kaluza--Klein Theory with spontaneous symmetry breaking and Higgs' mechanism.
A Lagrangian for fermions for geometrized bosonic part of GSW (Glashow--Salam--Weinberg) model in our approach
has been derived. Yukawa-type terms and mass terms coming from higher dimensions have been obtained.
In the paper 1/2-spin fields and 3/2-spin fields are considered.
\end{abstract}

\section{Introduction}
The aim of the paper is to develop a formalism of the \NK\ to include fermion fields getting \un\ of Yukawa coupling. One wants to develop
a formalism which is able to get masses for fermions from Yukawa coupling in such a way that Yukawa terms are coming from higher dimensions
(Kaluza--Klein) theory. Simultaneously we want to get chiral fermions. In order to do this one considers them as zero modes of the whole gauge group
manifolds. In this way one gets chiral fermions with masses from higher dimensions. A~novel approach consists
in using new kind of gauge \dv s in the \NK\ and an expansion of spinor fields in zero modes of the group manifold. In the case of the \NK\
with \ssb\ and \Hm\ we expand spinor fields in harmonics on $H=G/G_0$ or $S^2$ getting mass terms and Yukawa-type couplings. The use of zero modes
on a group manifold allows us to avoid Planck's mass terms for \fe s.

In the paper we consider fermions in the \NK. We suppose that \fe s belong to \fn\ \rp ation of generalized Lorentz groups
$\SO(1,n+3)$ ($\Spin(1,n+3)$), $\SO(1,n+n_1+3)$ ($\Spin(1,n+n_1+3)$) or $\SO(1,19)$ ($\Spin(1,19)$). We consider 0-form spinor fields
(i.e.\ spinor fields in the usual sense) and 1-form spinor fields. Due to this approach we get 1/2-spin fields and also 3/2-spin fields in our
theory.

The \E\nos\ Kaluza--Klein (Jordan--Thiry) Theory (a real version) has
been developed in the past (see Refs \cite1--\cite5). The theory unifies \gr
al theory described by NGT (\E\nos\ \E\gr al Theory, see Ref.~\cite6) and
\YM' fields (also \elm c field). In the case of the \JT\ this theory includes
scalar field. The \NK\ can be obtained from the \JT\ by simply putting this
scalar field to zero. In this way it is a limit of the \JT.

The \JT\ has several physical applications in \co y, e.g.: (1)~\co ical \ct,
(2)~inflation, (3)~quintessence, and some possible relations to the dark
matter problem. There is also a possibility to apply this theory to an
anomalous acceleration problem of Pioneer 10/11 (see Refs~\cite7, \cite{7a}).

Simultaneously the theory unifies gravity with gauge fields in a nontrivial
way via geometrical \un s of two fundamental \iv ce principles in Physics:
(1)~the \cd\ \iv ce \pp, (2)~the gauge \iv ce \pp. \E\un\ on the level of \iv
ce \pp s is more important and deeper
than on the level of interactions for from \iv ce
\pp s we get \cv\ laws (via the Noether theorem). In some sense Kaluza--Klein
theory unifies the energy-momentum \cv\ law with the \cv\ of a color
(isotopic) charge (an electric charge in an \elm c case).

This \un\ has been achieved in higher than four-\di al world, i.e.\
$(n+4)$-\di al, where $n=\dim G$, $G$~is a gauge group for a \YM' field,
which is a semisimple Lie group (\nA). In an \elm c case we have $G=\U(1)$ and
a \un\ is in \hbox{5-\di al} world (see also~\cite8). The \un\ has been achieved
via a natural \nos\ metrization of a fiber bundle. This metrization is
right-\iv t \wrt an action of a group~$G$. We present also an Hermitian
metrization of a fiber bundle in two versions: complex and \hc. The \cn\
on a fiber bundle of frames over a manifold~$P$ (a~bundle manifold) is
compatible with a metric tensor (\nos\ or Hermitian in complex or \hc\
version). In the case of $G=\U(1)$ the geometrical structure is bi\iv t
\wrt an action of~$\U(1)$, in a general \nA\ case this is only right-\iv t.

The \un\ is nontrivial for we
can get some additional effects unknown in conventional theories of gravity
and gauge fields (\YM' or \elm c field). All of these effects, which we call
\ti{interference effects} between gravity and gauge fields are testable in
\pp\ in experiment or in an observation. The formalism of this \un\ has been
described in Refs \cite1--\cite5,~\cite8 (without Hermitian versions).

It is possible to extend the \E\nos\ (\nA) Kaluza--Klein Theory to the case
of a spontaneous \s y breaking and \Hm\ (see Ref.~\cite1) by a nontrivial
combination of Kaluza \pp\ (Kaluza miracle) with \di al reduction procedure.
This consists in an extension of a base manifold of a principal fiber bundle
from $E$ (a~\spt) to $V=E\times M$, where $M=G/G_0$ is a manifold of
classical vacuum states. In the case of fermion fields it is possible to get
in the formalism Yukawa terms and due to a \ssb\ and \Hm\ masses for fermions.

The \NK\ is an example of the geometrization of \fn\ \ia\ (described by \YM'
and Higgs' fields) and \gr\ according to the Einstein program for
geometrization of \gr al and \elm c \ia s. It means an example to create a
Unified Field Theory. In the Einstein program we have to do with \elm sm and
gravity only. Now we have to do with more degrees of freedom, unknown in
Einstein times, i.e.\ GSW (Glashow--Salam--Weinberg) model, QCD, Higgs'
fields, GUT (Grand Unified Theories). Moreover, the program seems to be the
same.

We can paraphrase the definition from Ref.~\cite9: \ti{Unified Field Theory:
any theory which attemps to express \gr al theory and \fn\ \ia s theories
within a single unified framework. Usually an attempt to generalize
Einstein's general theory of relativity alone to a theory of gravity and
classical theories describing \fn\ \ia s}. In our case this single unified
framework is a multi\di al analogue of geometry from Einstein Unified Field
Theory (treated as generalized gravity) defined on principal fiber bundles
with base manifolds: $E$ or $E\times V$ and structural groups $G$ or~$H$.
Thus the definition from an old dictionary (paraphrased by us) is still valid.

Some ideas on geometrization and \un\ of fundamental physical \ia s can be found in Ref.~\cite K.

Summing up, \NK\ connects old ideas of unitary field theories (unified field
theories, see Refs \cite{10,11} for a review) with modern applications. This
is a geometrization and \un\ of a bosonic part of four \fn\ \ia s (see also Ref.~\cite{11a}).

\def\labelenumi{$\arabic{enumi}^\circ$}
Let us give some sketch of the development of the \E\nos\ \KK (Jordan--Thiry) Theory. In Ref.~\cite4 we introduced the theory in the \elm c case.
This case has been developed  in details in Ref.~\cite8. Ref.~\cite3 is devoted to a general \nA\ case. In Ref.~\cite1 one can find also \nA\
theory with \ssb\ and \Hm. The further development of \nA\ theory and \ssb\ with \Hm\ can be found in Ref.~\cite{11a}. The idea of a
dielectric model of color \cfn\ in the \NK\ has been introduced in Ref.~\cite2. This has been developed in Ref.~\cite{11a}. The further
development of a \s y breaking in the \NK\ (a~hierarchy of a \s y breaking) can be found in Ref.~\cite{beta}. Application of the scalar field
in the theory can be found in Refs \cite7, \cite{7a} for a Pioneer 10/11 anomaly acceleration. Some applications of this field in \co y can be
found in Ref.~\cite{gamma}. Ref.~\cite{gamma} gives some \co ical models with Higgs' field and a scalar field~$\Psi$ (as a quintessence field).
(Using the same notation~$\Psi$ for a scalar field in the \E\nos\ \KK (Jordan--Thiry) Theory and for spinor field~$\Psi$ or even for spinor
valued 1-form~$\Psi$ cannot cause any confusion.)
It deals in great details with the \co ical evolution of Higgs' field in the framework of \co ical models with this scalar field~$\Psi$ for
\E\nos\ \KK (Jordan--Thiry) Theory.
Section~2 of the paper is devoted to some details of the \NK\ with (and without) \ssb\ and \Hm. Section~3 gives many
details of application of the theory to the bosonic part of GSW (Glashow--Salam--Weinberg) model. The application achieves several important
issues:
\ben
\item mass spectrum of $W^\pm, Z^0$ bosons and Higgs' boson agreed with an experiment,
\item a correct value of Weinberg angle (also with radiative corrections).
\een
It is important to mention that in order to get a correct value for Higgs' boson mass we should consider complex Hermitian version of the theory.
Thus an experiment chooses the correct version of the theory among real, Hermitian (complex Hermitian, hypercomplex Hermitian).

Ref.~\cite{11a} contains all the details of these three approaches. In Ref.~\cite{11a} we give also some ideas to quantize the theory using
Yukawa--Efimov nonlocal quantization procedure within path integral quantization. Let us notice the following fact. One hundred years ago three men,
A.~Einstein, D.~Hilbert and O.~Klein were discussing what should be a \Lg\ for a \gr al field. They decided it should be a scalar \cvt\ for
a \LC \cn\ defined on a 4-\di al manifold (a~\spt). In that time an additional \Lg\ was known. It was a \Lg\ for an \elm c field $-\frac14
F_{\m}F^\m$ coming from Maxwell \e s which were known since about fifty years. Some years later T.~Kaluza designed a theory where $R_5$ (also
a~scalar \cvt) on 5-\di al manifold with additional \s ies is equal to $R_4-\frac{\la^2}4 F_\m F^\m$ ($\la=\frac{2\sqrt{G_N}}{c^2}$,
$G_N$~---~a~Newton \ct). Now, 100~years of General Relativity, almost
150~years of Maxwell \e s and almost 100~years of Kaluza's idea we came to the conclusion that only a scalar \cvt\ can be a~\Lg\ for unified
field theory of all physical \ia s. This will be a scalar \cvt\ of a \cn\ defined on many-\di al manifold with some \s ies and additional degrees of freedom
(e.g.\ skew-\s ic) metric. Moreover, we should consider an action for this scalar \cvt\ on a multi-\di al manifold as an integral.
In our approach this multi-\di al manifold is a bundle manifold and we integrate over the bundle manifold.

Let us notice that we geometrize Higgs' fields, \ssb\ and \Hm\ according to the Einstein program. Using our achievements from Ref.~\cite{11a}
we can obtain Yukawa-type terms in the \Lg\ for ordinary $\frac12$-spin fermions. In this way we are getting Yukawa couplings and masses for \fe s
from higher \di s.

Let us give the following remark on the \Lg\ in a field theory and mechanics. The \Lg\ is defined as a difference between kinetic and \pt\ energy.
Let us notice the following fact. In General Relativity a~scalar \cvt---a~\Lg\ for \gr al field---is a pure kinetic energy. The same is for
a~\Lg\ of a test \pc\ (a~\Lg\ in mechanics). This is also a pure kinetic energy. Further development considered in \KK (Jordan--Thiry) Theory
conserves this idea. In the \E\nos\ \KK (Jordan--Thiry) Theory we have the same in both cases:
\ben
\item for a field theory (geometrized and unified),
\item for a mechanics (a test \pc\ motion)---\gz\ \KWK \e, which is a geodetic \e\ obtained from \v al principle, where the \Lg\ is ``a~pure
kinetic energy'' on multi-\di al manifold.
\een
There is nothing wrong in a conventional \Hm\ (e.g.\ GSW model). Moreover, the \NK\ approach gives us:
\ben
\item Higgs' fields as gauge fields over ``a manifold of classical vacuum states'' ($S^2$~in the case of GSW model).
\item Proper spectrum of masses of $W^\pm,Z_0$ and Higgs' bosons, if a geometry on~$S^2$ is complex Hermitian (really K\"ahlerian).
\item \E\gz\ \KWK \e\ with some additional charges coupled Higgs' field to a test \pc\ (with a~possible test in an experiment).
\een
The theory is not longer a phenomenological theory. We are getting \Hm\ and masses for gauge bosons from higher \di s.

It is a geometrical \un\ of gravity and electro-weak \ia s as we described it above. In our meaning all \fn\ \ia s should be described by a
multi\di al linear \cn\ defined on a \nos ally metrized fiber bundle compatible with a metric \sc e. A~base manifold should contain a manifold
of classical vacuum states. This is a bosonic part of \ia.

Fermionic sector of our approach (our own method) should be described by a fermion field defined on the mentioned manifold and coupled to a~\cn\
defined on the same manifold (a~bosonic sector) in a minimal way. Our minimal coupling scheme (a~\ci\ multi\di al coupling) has been defined
in Refs \cite{ax4}, \cite{ax5}, \cite{ax6}, \cite{11a}, \cite{aQ}, \cite{r1} and developed here. This minimal coupling scheme couples effectively
multi\di al spinor field~$\Psi$ to a \cn\ which is metric (\wrt a \s ic part of a multi\di al \nos\ metric, the notion of a \nos\ metric is
of course an abuse of a nomination, moreover it is understable). In this way a coupling between spinor fields ($\frac12$-spin fermion fields) is
consistent as in 4-\di al case (see Refs \cite{q}, \cite{11a}, \cite{ax4}, \cite{ax5}). We consider $\Psi$ as zero modes on a group manifold
in order to avoid Planck's mass terms. We described also a minimal coupling scheme of a
$\frac32$-spin field for a future convenience using differential forms formalism.

For there is not any trace of super\s y and supergravity in an experiment we do not develop super\s y and supergravity approach even in
$\frac32$-spin fermions case. There is of course a place for such an approach in our method using supermanifolds with anticommuting
coordinates (see Refs \cite{omega}, \cite{rho}). Moreover, we will develop it elsewhere if the experiment gives us some important traces (future
accelerators).

In the paper we consider a \fe ic part of the \NK. It means, we couple spinor fields describing \fe s to a geometry on the
Kaluza--Klein manifold. We suppose that our spinors are coupled to \LC \cn\ defined on \KK manifold generated by a \s ic part of a metric tensor
$\g_\(AB)$ or $\g_\(\tA\tB)$. In this way we want to get a \un\ of gravity and \YM' fields, Higgs' fields coupled to \fe s. Up to now our \un\
describes only a bosonic part of the theory. We work in the following way.
We couple spinors to \nA\ \KK Theory via new kind of gauge \dv s. These \dv s induce on many-\di al
manifold a new \cn, which is metric but with a non-vanishing torsion. In this way we get a consistency between differentiation of spinors and
tensors (vectors). We couple also our spinors (many-\di al) on an extended manifold with Higgs' fields also (not only 4-\di al \YM' fields) and
we give a comprehensive treatment of a spinor coupling to a bosonic part of GSW (Glashow--Salam--Weinberg) model in our approach. We get a \un\
of Yukawa coupling e.g.\ in GSW model. We suppose that
our spinor fields are zero-modes of $\Dint$ operator. In this way we remove very heavy \fe s (with masses of the Planck's mass order) from the
theory. We develop these \fe\ fields in zero-modes \f s and also in generalized harmonics on a compact manifold $M=G/G_0$. In the case of
GSW-model we have $S^2$ and of course spherical harmonics.
We get several interesting terms in the \Lg\ for \fe s and we interpret them. We get several terms coming to mass terms and also to some mixings.
Eventually we proceed a \s y breaking from $\SO(1,n+3)$ ($\Spin(1,n+3)$), $\SO(1,n+n_1+3)$ ($\Spin(1,n+n_1+3)$) or $\SO(1,19)$ ($\Spin(1,19)$) to
a Lorentz group $\SO(1,3)$ (${\rm SL}(2,\C)$) getting a tower of spinor fields ($0$-forms and $1$-forms). Spinors (zero modes) are coupled by new
gauge \dv s. This is a new point.

Let us give some remarks on a coupling between Rarita--Schwinger field ($\frac32$-spin fermion field) and gravity. For a very long time a~coupling
between gravity and $\frac32$-spin fields was badly defined. In \cite{a}, \cite{bb} a new coupling was introduced between gravity and
Rarita--Schwinger fields, which solved the problem of coupling to gravity. It was obtained by two independent ways. In super-\gr\ theory~\cite{a},
and by using the formalism of differential forms~\cite{bb}. We are using the second approach and extending it to the \KK theory. The minimal
coupling for Rarita--Schwinger fields to an \elm c field (in general, to any gauge field)
is inconsistent for we have the Velo--Zwanziger paradox~\cite{c} in the scheme. The Rarita--Schwinger \e\
is relativistically \ci\ but solutions can be acausal. This happens due to algebraic constraints which are differential consequences of the
Rarita--Schwinger \e\ \cite{c},~\cite{d}. These constraints depend on the strength of the \elm c field. In the paper \cite{e} we consider new
kinds of gauge \dv s for spin-$\frac32$ fields (see \cite{ax4}, \cite{ax5}). In such a way we generalize the minimal coupling scheme. We use
differential forms for Rarita--Schwinger fields as in~\cite{bb}. But we define spin-$\frac32$ fields on the \KK manifold. This is of course
a~5-\di al case. From this we get a new term in the \Lg. This term describes the \ia\ between the \elm c field and a dipole electric moment
of the $\frac32$-spinor field of value about $10^{-31}\,{\rm[cm]}\,q$ (see Ref.~\cite{e}). Thus this term is very small. The term is an
``interference'' effect between the \gr al and \elm c fields. It breaks PC and~T and is similar to a~term given in Refs \cite{ax4},
\cite{ax5},~\cite{aQ}. Our new term has an important consequence. Namely, it implies that the first differential consequences for the
Rarita--Schwinger \e\ (obtained from this \Lg) are differential \e s. We do not obtain algebraic constraints involving \elm c field. Thus the
Velo--Zwanziger paradox is avoided. In Ref.~\cite{w} one gets also a nonminimal coupling between the Rarita--Schwinger field and an \elm c field.
The additional term in Ref.~\cite{w} does not violate the PC \s y and is much different from our term. The most \fn\ difference between $N=2$
supergravity approach (see Ref.~\cite{w}) and our approach (see Ref.~\cite{e}) consists not only in the PC breaking, but in the fact that our
Rarita--Schwinger field is \ti{charged\/} and Rarita--Schwinger field in Ref.~\cite{w} is \ti{uncharged}. In this way we have the ordinary coupling
between the \elm c field and the Rarita--Schwinger field plus new term, and we avoid the Velo--Zwanziger paradox. In~\cite{w} this does not occur,
because Rarita--Schwinger field is \ti{uncharged\/} and the paradox does not take place. However, we pay a price for avoiding the Velo--Zwanziger
paradox. Our theory breaks~PC. Fortunately this breaking is very small. We point out that our theory is not a supergravity-like theory, and we
do not use anticommuting Majorana spinors.

Afterwards we consequently develop the formalism in the \nA\ \KK Theory. The coupling between $\frac32$-spin field is consistently using
differential forms formalism (see Ref.~\cite{aQ}). This coupling (in Ref.~\cite{aQ}) is a \gn\ of a minimal coupling scheme and is going to some
kind of PC-breaking terms (of the same order as in 5-\di al case). In the paper we extend the approach to the case with \ssb\ and \Hm. Moreover,
the paper is devoted basically to $\frac12$-spin fermion fields. All the fields considered here are zero modes of the group
manifold, which is different from previous approaches (see Refs \cite{aQ}, \cite{r3}, \cite{alfa}). This is supposed in order to avoid Planck's
mass terms in the \Lg\ for fermion fields. A~detailed formalism for $\frac32$-spin fermion fields will be developed elsewhere as we mention in the
text. We expect to avoid the Velo--Zwanziger paradox also in the case with \ssb\ and \Hm, getting also Yukawa-type terms. The absence of
Velo--Zwanziger paradox is quite obvious because differential consequences of the field \e\ for $\frac32$-spin \fe\ field are here \ti{differential
\e s}, \ti{not algebraic constraints}.

The paper is divided into five sections. In the second section we describe the \NK\ in general \nA\ case and also with a \sn\ \s y breaking taken
into account. In the third section we describe a bosonic part of GSW-model in the theory. In the fourth section we deal with spinor fields on
a~manifold~$P$. The fifth section is devoted to the \Lg\ of \fe s in the theory. We consider and discuss several possibilities of such a \Lg. We
find several interesting terms in \Lg s. We develop (as we mentioned before) spinor fields into a \gz\ Fourier series of \gz\ harmonics and
zero-modes \f\ of~$\Dint$. We consider also the problem of chiral \fe s using arguments of Atiyah--Singer index theorem. Eventually we proceed a
\s y breaking to the Lorentz group getting a tower of spinor fields. This is a new point in the \NK.

In Appendix A we give some \el s of Clifford algebras and Dirac matrices applicable for our theory. In Appendix~B we consider \ci\ \dv s in our
theory. Appendix C is devoted to Atiyah--Singer index theorem applicable for $\Dint$ elliptic operator defined on a compact group $G$ or~$H$. In
Conclusions we give some prospects for further research.

\section{Elements of the \NK\ in general \nA\ case and with \sn\ \s y breaking and \Hm}\label{s:els}
Let $\ul P$ be a principal fiber bundle over a \spt\ $E$ with a structural
group~$G$ which is a semisimple Lie group. On a \spt~$E$ we define a \nos\
tensor $g_\m=g_\(\m)+g_\[\m]$ \st
\beq2.1
\bal
g&=\det(g_\m)\ne0\\
\wt g&=\det(g_\(\m))\ne0.
\eal
\e
$g_\[\m]$ is called as usual a skewon field (e.g.\ in NGT, see Refs
\cite{6,8}).
We define on $E$ a \nos\ \cn\ compatible with $g_\m$ \st
\beq2.2
\ov Dg_{\a\b}=g_{\a\d}\gd \ov Q,\d,\b\g,(\ov\G)\ov \t{}^\g
\e
where $\ov D$ is an exterior covariant \dv\ for a \cn\ $\gd\ov\o,\a,\b,=
\gd\ov \G,\a,\b\g,\ov \t{}^\g$ and $\gd\ov Q,\a,\b\d,$ is its torsion. We suppose also
\beq2.3
\gd\ov Q,\a,\b\a,(\ov \G)=0.
\e
We introduce on $E$ a second \cn
\beq2.4
\gd\ov W,\a,\b,=\gd\ov W,\a,\b\g,\ov \t{}^\g
\e
\st
\bg2.5
\gd \ov W,\a,\b,=\gd \ov\o,\a,\b,-\tfrac23\,\gd \d,\a,\b,\ov W\\
\ov W=\ov W_\g\ov \t{}^\g=\tfrac12(\gd \ov W,\si,\g\si,-\gd \ov W,\si,\si\g,)
\ov \t{}^\g. \lb2.6
\e

Now we turn to \nos\ metrization of a bundle $\ul P$. We define a \nos\ tensor
$\g$ on a bundle manifold $P$ \st
\beq2.6a
\g=\pi^* g\op \ell_{ab}\t^\a\ot \t^b
\e
where $\pi$ is a projection from $P$ to $E$. On $\ul P$ we define a \cn~$\o$
(a~1-form with values in a Lie algebra $\fg$ of~$G$). In this way we can
introduce on~$P$ (a~bundle manifold) a frame  $\t^A=(\pi^*(\ov \t{}^\a),
\t^a)$ \st
$$
\t^a=\la\o^a,\q \o=\o^a X_a, \q a=5,6,\dots,n+4, \q
n=\dim G=\dim\fg, \q \la={\rm const.}
$$
Thus our \nos\ tensor looks like
\bg2.7
\g=\g_{AB} \t^A\ot\t^B, \q A,B=1,2,\dots,n+4,\\
\ell_{ab}=h_{ab}+\mu k_{ab}, \lb2.8
\e
where $h_{ab}$ is a bi\iv t Killing--Cartan tensor on~$G$ and $k_{ab}$ is a
right-\iv t skew-\s ic tensor on~$G$, $\mu={\rm const}$.

We have
\beq2.9
\bal
h_{ab}&=\gd C,c,ad,\gd C,d,bc,=h_{ab}\\
k_{ab}&=-k_{ba}
\eal
\e
Thus we can write
\bea2.10
\ov \g(X,Y)=\ov g(\pi'X,\pi'Y)+\la^2h(\o(X),\o(Y))\\
\ul \g(X,Y)=\ul g(\pi'X,\pi'Y)+\la^2k(\o(X),\o(Y)) \lb2.11
\e
($\gd C,a,bc,$ are structural \ct s of the Lie algebra $\fg$).

$\ov \g$ is the \s ic part of $\g$ and $\ul \g$ is the anti\s ic part of~$\g$.
We have as usual
\beq2.12
[X_a,X_b]=\gd C,c,ab,X_c
\e
and
\beq2.13
\O=\frac12 \gd H,a,\m,\t^\mu \land \t^\nu
\e
is a curvature of the \cn\ $\o$,
\beq2.14
\O=d\o+\frac12[\o,\o].
\e
The frame $\t^A$ on $P$ is partially nonholonomic. We have
\beq2.15
d\t^a=\frac\la2\Bigl(\gd H,a,\m,\t^\mu\land \t^\nu - \frac1{\la^2}\,
\gd C,a,bc,\t^b\land\t^c\Bigr)\ne0
\e
even if the bundle $\ul P$ is trivial, i.e.\ for $\O=0$. This is different than
in an \elm c case (see Ref.~\cite3). Our \nos\ metrization of a
principal fiber bundle gives us a right-\iv t structure on~$P$ \wrt an action
of a group~$G$ on~$P$ (see Ref.~\cite3 for more details). Having $P$ \nos
ally metrized one defines two \cn s on~$P$ right-\iv t \wrt an action of a
group $G$ on~$P$. We have
\beq2.16
\g_{AB}=\left(\begin{array}{c|c}
g_{\a\b}&0\\
\hline
0&\ell_{ab}\end{array}\right)
\e
in our left horizontal frame $\t^A$.
\bg2.17
D\g_{AB}=\g_{AD}\gd Q,D,BC,(\G)\t^C\\
\gd Q,D,BD,(\G)=0 \label{2.18}
\e
where $D$ is an exterior covariant \dv\ \wrt a \cn\ $\gd \o,A,B,=\gd \G,A,BC,
\t^C$ on~$P$ and $\gd Q,A,BC,(\G)$ its torsion. One can solve
Eqs~\er{2.17}--\er{2.18} getting the following results
\beq2.19
\gd \o,A,B,=\left(
\begin{array}{c|c}
\pi^*(\gd \ov\o,\a,\b,)-\ell_{db}g^{\mu\a}\gd L,d,\mu\b,\t^b&\gd L,a,\b\g,\t^\g\\
\hline
\ell_{bd}g^{\a\b}(2\gd H,d,\g\b,-\gd L,d,\g\b,)\t^\g & \gd \wt\o,a,b,
\end{array}\right)
\e
where $g^{\mu\a}$  is an inverse tensor of $g_{\a\b}$
\beq2.20
g_{\a\b}g^{\g\b}=g_{\b\a}g^{\b\g}=\d^\g_\a,
\e
$\gd L,d,\g\b,=-\gd L,a,\b\g,$ is an Ad-type tensor on~$P$ \st
\beq2.21
\ell_{dc}g_{\mu\b}g^{\g\mu}\gd L,d,\g\a,+\ell_{cd}g_{\a\mu}g^{\mu\g}
\gd L,d,\b\g,=2\ell_{cd}g_{\a\mu}g^{\mu\g}\gd H,d,\b\g,,
\e
$\gd\wt\o,a,b,=\gd \wt\G,a,bc,\t^c$ is a \cn\ on an internal space (typical
fiber) compatible with a metric $\ell_{ab}$ \st
\bg2.22
\ell_{db}\gd \wt\G,d,ac,+\ell_{ad}\gd \wt\G,d,cb,=-\ell_{db}\gd C,d,ac,\\
\gd \wt\G,a,ba,=0, \q \gd \wt\G,a,bc,=\gd -\wt\G,a,cb, \lb2.23
\e
and of course $\gd \wt Q,a,ba,(\wt\G)=0$ where $\gd \wt Q,a,bc,(\G)$ is a
torsion of the \cn~$\gd \wt\o,a,b,$.

We also introduce an inverse tensor of $g_\(\a\b)$
\beq2.24
g_\(\a\b)\wt g{}^\(\a\g)=\d^\g_\b.
\e
We introduce a second \cn\ on~$P$ defined as
\beq2.25
\gd W,A,B,=\gd \o,A,B,-\frac4{3(n+2)}\,\gd\d,A,B,\ov W.
\e
$\ov W$ is a horizontal one form
\bg2.26
\ov W=\hor\ov W\\
\ov W=\ov W_\nu\t^\nu =\tfrac12(\gd\ov W,\si,\nu\si, - \gd \ov W,\si,\si\nu,).
\lb2.27
\e

In this way we define on $P$ all analogues of four-\di al quantities from NGT
(see Refs \cite{6,14,15,16}). It means, $(n+4)$-\di al analogues from Moffat
theory of \gr, i.e.\ two \cn s and a \nos\ metric $\g_{AB}$. Those quantities
are right-\iv t \wrt an action of a group~$G$ on~$P$. One can calculate a
scalar curvature of a \cn\ $\gd W,A,B,$ getting the following result (see
Refs \cite{1,3}):
\beq2.28
R(W)=\ov R(\ov W)-\frac{\la^2}4 \bigl(2\ell_{cd}H^cH^d - \ell_{cd}L^{c\m}
\gd H,d,\m,\bigr) + \wt R(\wt \G)
\e
where
\beq2.29
R(W)=\g^{AB}\bigl(\gd R,C,ABC,(W)+\tfrac12 \, \gd R,C,CAB,(W)\bigr)
\e
is a Moffat--Ricci curvature scalar for the \cn~$\gd W,A,B,$,
$\ov R(\ov W)$ is a Moffat--Ricci curvature scalar for the \cn~$\gd \ov W,\a,\b,$,
and $\wt R(\wt \G)$ is a Moffat--Ricci curvature scalar for the \cn~$\gd \wt\o,a,b,$,
\bg2.30
H^a=g^\[\m]\gd H,a,\m,\\
L^{a\m}=g^{\a\mu}g^{\b\nu}\gd L,a,\a\b,. \lb2.31
\e
Usually in ordinary (\s ic) Kaluza--Klein Theory one has
$\la=2\frac{\sqrt{G_N}}{c^2}$, where $G_N$ is a Newtonian \gr al \ct\ and
$c$~is the speed of light. In our system of units $G_N=c=1$ and $\la=2$. This
is the same as in \NK\ in an \elm ic case (see Refs \cite{4,8}). In the \nA\
Kaluza--Klein Theory which unifies GR and \YM\ field theory we have a \YM\
lagrangian and a \co ical term. Here we have
\beq2.32
\cLY=-\frac1{8\pi}\,\ell_{cd}\bigl(2H^cH^d-L^{c\m}\gd H,d,\m,\bigr)
\e
and $\wt R(\wt\G)$ plays a role of a \co ical term.

In order to incorporate a \sn\ \s y breaking and \Hm\ in our geometrical \un\
of \gr\ and \YM' fields we consider a fiber bundle $\ul P$ over a base manifold
$E\times G/G_0$, where $E$ is a \spt, $G_0\subset G$, $G_0,G$ are semisimple Lie
groups. Thus we are going to combine a Kaluza--Klein theory with a \di al
reduction procedure.

Let $\ul P$ be a principal fiber bundle over $V=E\times M$ with a structural
group~$H$ and with a projection~$\pi$, where $M=G/G_0$ is a homogeneous
space, $G$~is a semisimple Lie group and $G_0$ its semisimple Lie subgroup.
Let us suppose that $(V,\g)$ is a manifold with a \nos\ metric tensor
\beq3.1
\g_{AB}=\g_\(AB)+\g_\[AB].
\e
The signature of the tensor $\g$ is ${(}+{-}-{-},
\underbrace{{}-{-}-\cdots-}_{n_1})$. Let us
introduce a natural frame on~$\ul P$
\beq3.2
\t^{\tilde A}=(\pi^*(\t^A),\t^0=\la\o^a), \q \la={\rm const.}
\e
It is convenient to introduce the following notation. Capital Latin indices
with tilde $\wt A,\wt B,\wt C$ run $1,2,3,\dots,m+4$, $m=\dim H+\dim M
=n+\dim M=n+n_1$, $n_1=\dim M$, $n=\dim H$. Lower Greek indices $\a,\b,\g,\d
=1,2,3,4$ and lower Latin indices
$a,b,c,d=n_1+5,n_2+5,\dots,\break n_1+6,\dots,m+4$. Capital Latin indices
$A,B,C=1,2,\dots,n_1+4$. Lower Latin indices with tilde $\wt a,\wt b,\wt c$
run $5,6,\dots,n_1+4$. The symbol ``$\ov{\phantom{N}}$'' over $\t^A$ and other quantities indicates
that these quantities are defined on~$V$. We have of course
$$
n_1=\dim G-\dim G_0=n_2-(n_2-n_1),
$$
where $\dim G=n_2$, $\dim G_0=n_2-n_1$, $m=n_1+n$.

On the group $H$ we define a bi-\iv t (\s ic) Killing--Cartan tensor
\beq3.3
h(A,B)=h_{ab}A^aB^b.
\e
We suppose $H$ is semisimple, it means $\det(h_{ab})\ne0$. We define a skew-\s
ic right-\iv t tensor on~$H$
$$
k(A,B)=k_{bc}A^bB^c, \q k_{bc}=-k_{cb}.
$$

Let us turn to the \nos\ metrization of~$\ul P$.
\beq3.4
\ka(X,Y)=\g(X,Y)+\la^2\ell_{ab}\o^a(X)\o^b(Y)
\e
where
\beq3.5
\ell_{ab}=h_{ab}+\xi k_{ab}
\e
is a \nos\ right-\iv t tensor on~$H$. One gets in a matrix form (in the
natural frame \er{3.2})
\beq3.6
\ka_{\tA\tB}=\left(\begin{array}{c|c}
\g_{AB}&0\\ \hline 0&\ell_{ab}
\end{array}\right),
\e
$\det(\ell_{ab})\ne0$, $\xi={\rm const}$ and real, then
\beq3.7
\ell_{ab}\ell^{ac}=\ell_{ba}\ell^{ca}=\gd\d,c,b,.
\e
The signature of the tensor $\ka$ is $(+,-{-}-,\underbrace{-\cdots-}_{n_1},
\underbrace{{}-{-}\cdots-}_n)$.
As usual, we have commutation relations for Lie algebra of~$H$, $\fh$
\beq3.8
[X_a,X_b]=\gd C,c,ab,X_c.
\e
This metrization of $\ul P$ is right-\iv t \wrt an action of $H$ on $P$.

Now we should \nos ally metrize $M=G/G_0$. $M$~is a homogeneous space for~$G$
(with left action of group~$G$). Let us suppose that the Lie algebra of~$G$,
$\fg$ has the following reductive decomposition
\beq3.9
\fg=\fg_0 \mathrel{\dot+} \fm
\e
where $\fg_0$ is a Lie algebra of $G_0$ (a~subalgebra of~$\fg$) and $\fm$
(the complement to the subalgebra~$\fg_0$) is $\Ad G_0$ \iv t, $\dot+$
means a direct sum. Such a decomposition might be not unique, but we assume
that one has been chosen. Sometimes one assumes a stronger condition
for~$\fm$, the so called \s y requirement,
\beq3.10
[\fm,\fm]\subset \fg_0.
\e
Let us introduce the following notation for generators of $\fg$:
\beq3.11
Y_i\in\fg, \q Y_{\tilde\imath}\in\fm, \q Y_{\hat a}\in \fg_0.
\e
This is a de\cm\ of a basis of $\fg$ according to \er{3.9}. We define a \s ic
metric on~$M$ using a \KC form on~$G$ in a classical way. We call this tensor
$h_0$.

Let us define a tensor field $h^0(x)$ on $G/G_0$, $x\in G/G_0$, using tensor
field $h$ on~$G$. Moreover, if we suppose that $h$ is a bi\iv t metric on~$G$
(a~\KC tensor) we have a simpler construction.

The complement $\fm$ is a tangent space to the point $\{\ve G_0\}$ of~$M$,
$\ve$~is a unit element of~$G$. We restrict $h$ to the space $\fm$ only. Thus
we have $h^0(\{\ve G_0\})$ at one point of~$M$. Now we propagate
$h^0(\{fG_0\})$ using a left action of the group $G$
$$
h^0(\{fG_0\})=(L_f^{-1})^\ast (h^0(\{\ve G_0\})).
$$
$h^0(\{\ve G_0\})$ is of course $\Ad G_0$ \iv t tensor defined on~$\fm$
and $L_f^\ast h^0=h^0$.

We define on $M$ a skew-\s ic 2-form $k^0$. Now we introduce a natural frame
on~$M$. Let $\gd f,i,jk,$ be structure \ct s of the Lie algebra $\fg$, i.e.
\beq3.12
[Y_j,Y_k]=\gd f,i,jk, Y_i.
\e
$Y_j$ are generators of the \Li~$\fg$. Let us take a local section $\si:V\to
G/G_0$ of a natural bundle $G\mapsto G/G_0$ where $V\subset M=G/G_0$. The
local section $\si$ can be considered as an introduction of a \cd\ system
on~$M$.

Let $\o_{MC}$ be a left-\iv t Maurer--Cartan form and let
\beq3.13
\gd \o,\si,MC,=\si^\ast \o_{MC}.
\e
Using de\cm\ \er{3.9} we have
\beq3.14
\gd \o,\si,MC,=\gd \o,\si,0,+\gd \o,\si,\fm,=
\t^{\hi}Y_{\hi}+\ov t{}^{\td a}Y_{\td a}.
\e
It is easy to see that $\ov \t{}^{\td a}$ is the natural (left-\iv t) frame
on~$M$ and we have
\bea3.15
h^0&=\gd h,0,\td a\td b,\ov\t{}^{\td a}\otimes \ov\t{}^{\td b}\\
k^0&=\gd k,0,\td a\td b,\ov\t{}^{\td a}\land \ov\t{}^{\td b}. \lb3.16
\e
According to our notation $\wt a,\wt b=5,6,\dots,n_1+4$.

Thus we have a \nos\ metric on $M$
\beq3.17
\g_{\td a\td b}=r^2\bigl(\gd h,0,\td a\td b,+\z \gd k,0,\td a\td b,\bigr)
=r^2g_{\td a\td b}.
\e
Thus we are able to write down the \nos\ metric on $V=E\tm M=E\tm G/G_0$
\beq3.18
\g_{AB}=\left(\begin{array}{c|c}
g_{\a\b}&0\\
\hline
0 & r^2g_{\td a\td b}
\end{array}\right)
\e
where
\begin{align*}
g_{\a\b}&=g_\(\a\b)+g_\[\a\b]\\
g_{\td a\td b}&=\gd h,0,\td a\td b,+\z \gd k,0,\td a\td b,\\
\gd k,0,\td a\td b,&=-\gd k,0,\td b\td a,\\
\gd h,0,\td a\td b,&=\gd h,0,\td b\td a,,
\end{align*}
$\a,\b=1,2,3,4$, $\wt a,\wt b=5,6,\dots,n_1+4=\dim M+4=\dim G-\dim G_0+4$.
The frame $\ov\t{}^{\td a}$ is unholonomic:
\beq3.19
d\ov\t{}^{\td a}=\frac12\,\gd \ka,\td a,\td b\td c, \ov\t{}^{\td b}\land
\ov \t{}^{\td c}
\e
where $\gd \ka,\td a,\td b\td c,$ are \cf s of nonholonomicity and depend  on
the point of the manifold $M=G/G_0$ (they are not \ct\ in general). They
depend on the section~$\si$ and on the \ct s $\gd f,\td a,\td b\td c,$.

We have here three groups $H,G,G_0$. Let us suppose that there exists a
homomorphism $\mu$ between $G_0$ and~$H$,
\beq3.20
\mu:G_0 \to H
\e
\st a centralizer of $\mu(G_0)$ in $H$, $C^\mu$ is isomorphic to~$G$.
$C^\mu$, a centralizer of $\mu(G_0)$ in $H$, is a set of all \el s of~$H$
which commute with \el s of $\mu(G_0)$, which is a subgroup of~$H$. This
means that $H$~has the following structure, $C^\mu=G$.
\beq3.21
\mu(G_0)\otimes G\subset H.
\e
If $\mu$ is a iso\mo sm between $G_0$ and $\mu(G_0)$ one gets
\beq3.22
G_0\otimes G\subset H.
\e
Let us denote by $\mu'$ a tangent map to $\mu$ at a unit \el. Thus $\mu'$ is
a differential of~$\mu$ acting on the \Li\ \el s. Let us suppose that the
\cn~$\o$ on the fiber bundle $\ul P$ is \iv t under group action of~$G$ on the
manifold $V=E\tm G/G_0$. According to Refs \cite{13,18,19,20} this means the
following.

Let $e$ be a local section of $\ul P$, $e:V\subset U\to P$ and $A=e^*\o$. Then
for every $g\in G$ there exists a gauge \tf\ $\rho_g$ \st
\beq3.23
f^*(g)A=\Ad_{\rho_g^{-1}}A+\rho_g^{-1}\,dg_g,
\e
$f^*$ means a pull-back of the action $f$ of the group $G$ on the
manifold~$V$. According to Refs \cite{18,19,20,21,22,23}
(see also Refs \cite{**,*1,*2}) we are able to write
a general form for such an~$\o$. Following Ref.~\cite{20} we have
\beq3.24
\o=\wt \o_E +\mu'\circ \gd \o,\si,0,+\F\circ \gd \o,\si,\fm,.
\e
(An action of a group $G$ on $V=E\tm G/G_0$ means left multiplication on a
homogeneous space $M=G/G_0$.)
where $\gd \o,\si,0,+\gd \o,\si,\fm,=\gd \o,\si,MC,$ are components of the
pull-back of the Maurer--Cartan form from the de\cm~\er{3.14}, $\wt\o_E$ is a
\cn\ defined on a fiber bundle $Q$ over a \spt~$E$ with structural
group~$C^\mu$ and a projection~$\pi_E$. Moreover, $C^\mu=G$ and  $\wt\o_E$ is
a 1-form with values in the \Li~$\fg$. This \cn\ describes an ordinary \YM'
field gauge group $G=C^\mu$ on the \spt~$E$. $\F$~is a \f\ on~$E$ with values
in the space $\wt S$ of linear maps
\beq3.25
\F:\fm \to \fh
\e
\sf ying
\beq3.26
\F([X_0,X])=[\mu'X_0,\F(X)], \q X_0\in\fg_0.
\e
Thus
\beq3.27
\bal
\wt\o_E=\gd\wt\o,i,E,Y_i, &\q Y_i\in\fg,\\
\gd\o,\si,0,=\t^{\hi}Y_{\hi}, &\q Y_{\hi}\in\fg_0,\\
\gd\o,\si,\fm,=\ov\t{}^{\td a}Y_{\td a}, &\q Y_{\td a}\in\fm.
\eal
\e

Let us write condition \er{3.24} in the base of left-\iv t form
$\t^{\hi},\ov\t{}^{\td a}$, which span respectively dual spaces
to~$\fg_0$ and~$\fm$ (see Refs \cite{24,25}). It is easy to see that
\beq3.28
\F\circ \gd\o,\si,\fm,=\gd \F,a,\td a,(x)\ov\t{}^{\td a}X_a, \q X_a\in\fh
\e
and
\beq3.29
\mu'=\gd\mu,a,\hi, \t^{\hi}X_a.
\e
From \er{3.26} one gets
\beq3.30
\F_{\td b}^c (x)f_{\hi\td a}^{\td b}=
\mu^a_{\hi}\F^b_{\td a}(x)\gd C,c,ab,
\e
where $f^{\td b}_{\hi\td a}$ are structure \ct s of the \Li~$\fg$ and
$\gd C,c,ab,$ are structure \ct s of the \Li~$\fh$. Eq.~\er{3.30} is a
constraint on the scalar field $\F^a_{\td a}(x)$. For a curvature of~$\o$ one
gets
\bml3.30a
\O=\frac12\,\gd H,C,AB,\t^A\land \t^BX_C=
\frac12\,\gd\wt H,i,\m,\t^\mu\land\t^\nu \a^c_iX_c
+\gv{\n_\mu} \F^c_{\td a}\t^\mu\land \t^{\td a}X_c\\
{}+\frac12\,\gd C,c,ab,\F^a_{\td a}\F^b_{\td b}\t^{\td a}\land \t^{\td b}
X_c - \frac12\,\F^c_{\td d}f^{\td d}_{\td a\td b}\t^{\td a}\land\t^{\td b}X_c
-\mu^c_{\hi} \gd f,\hi,\td a\td b,\t^{\td a}\land\t^{\td b}X_c.
\e
Thus we have
\bg3.31
\gd H,c,\m,=\gd \a,c,i,\gd \wt H,i,\m,\\
\gd H,c,\mu\td a,=\gv{\n_\mu} \F^c_{\td a}=-\gd H,c,\td a\mu, \lb3.32 \\
\gd H,c,\td a\td b,=\gd C,c,ab,\cdot\F^a_{\td a}\F^b_{\td b} - \mu^c_{\hi}
f^{\hi}_{\td a\td b} - \F^c_{\td d}\gd f,\td d,\td a\td b, \lb3.33
\e
where $\gv{\n_\mu}$ means gauge \dv\ \wrt the \cn\ $\wt\o_E$ defined on a
bundle~$Q$ over a \spt~$E$ with a \sc al group~$G$
\beq3.34
Y_i=\a^c_i X_c.
\e
$\gd \wt H,i,\m,$ is the curvature of the \cn\ $\ov\o_E$ in the base
$\{Y_i\}$, generators of the \Li\ of the Lie group~$G$, $\fg$, $\a^c_i$ is
the matrix which connects $\{Y_i\}$ with $\{X_c\}$. Now we would like to
remind that indices $a,b,c$ refer to the \Li~$\fh$, $\wt a,\wt b,\wt c$ to
the space~$\fm$ (tangent space to~$M$), $\wh \imath,\wh \jmath,\wh k$ to the
\Li~$\fg_0$ and $i,j,k$ to the \Li\ of the group $G$, $\fg$. The matrix
$\a^c_i$ establishes a direct relation between generators of the \Li\ of the
subgroup of the group~$H$ iso\mo c to the group~$G$.

Let us come back to a construction of the \NK\ on a manifold~$P$. We should
define \cn s. First of all, we should define a \cn\ compatible with a \nos\
tensor $\g_{AB}$, Eq.~\er{3.18},
\bg3.35
\gd \ov\o,A,B,=\gd \ov\G,A,BC,\t^C\\
\ov D\g_{AB}=\g_{AD}\gd\ov Q,D,BC,(\ov\G)\t^C \lb3.36 \\
\gd \ov Q,D,BD,(\ov \G)=0 \nn
\e
where $\ov D$ is the exterior covariant \dv\ \wrt $\gd \ov\o,A,B,$ and $\gd
\ov Q,D,BC,(\ov\G)$ its torsion.

Using \er{3.18} one easily finds that the \cn\ \er{3.35} has the following
shape
\beq3.37
\gd \ov\o,A,B,=\left(\begin{array}{c|c}
\pi^*_E(\gd \ov\o,\a,\b,) & 0 \\ \hline
0 & \gd \wh{\bar\o},\td a,\td b,
\end{array}\right)
\e
where $\gd \ov\o,\a,\b,=\gd \ov \G,\a,\b\g,\ov\t{}^\g$ is a \cn\ on the
\spt~$E$ and $\gd\wh{\ov \o},\td a,\td b,=\gd \wh{\ov\G},\td a,\td b\td c,
\ov \t{}^{\td c}$ on the manifold $M=G/G_0$ with the following properties
\bg3.38
\ov Dg_{\a\b}=g_{\a\d}\gd \ov Q,\d,\b\g,(\ov \G)\ov \t{}^\g=0\\
\gd \ov Q,\a,\b\a,(\ov\G)=0 \lb3.39 \\
\wh{\ov D}g_{\td a\td b}=g_{\td a\td d}\gd\wh{\ov Q},\td d,\td b\td c,(\wh{\ov\G}).
\lb3.40 \\
\gd\wh{\ov Q},\td d,\td b\td d,(\wh{\ov\G})=0 \nonumber
\e

$\ov D$ is an exterior covariant \dv\ \wrt a \cn~$\gd \ov\o,\a,\b,$. $\gd \ov
Q,\a,\b\g,$ is a tensor of torsion of a \cn~$\gd \ov\o,\a,\b,$. $\wh{\ov
D}$~is an exterior covariant \dv\ of a \cn~$\gd \wh{\ov\o},\td a,\td b,$ and
$\gd \wh{\ov Q},\td a,\td b\td c,(\wh{\ov\G})$ its torsion.

On a \spt\ $E$ we also define the second affine \cn\ $\gd \ov W,\a,\b,$ \st
\beq3.41
\gd \ov W,\a,\b,= \gd \ov\o,\a,\b, - \frac23\,\gd \d,\a,\b,\ov W,
\e
where
$$
\ov W=\ov W_\g \ov\t{}^\g = \tfrac12(\gd \ov W,\si,\g\si,-\gd \ov W,\si,\g\si,).
$$
We proceed a \nos\ metrization of a principal fiber bundle $\ul P$ according
to \er{3.18}. Thus we define a right-\iv t \cn\ \wrt an action of the
group~$H$ compatible with a tensor $\ka_{\tA\tB}$
\bg3.42
D\ka_{\tA\tB}=\ka_{\tA\td D}\gd Q,\td D,\tB\tC,(\G)\t^{\tC}\\
\gd Q,\td D,\td B\td D,(\G)=0 \nn
\e
where $\gd \o,\tA,\tB,=\gd \G,\tA,\tB\tC,\wt \t{}^\tC$. $D$ is an exterior
covariant \dv\ \wrt the \cn\ $\gd \o,\tA,\tB,$ and $\gd Q,\tA,\tB\tC,$ its
torsion. After some calculations one finds
\beq3.43
\gd \o,\tA,\tB,= \bma
\pi^*(\gd \ov\o,A,B,)-\ell_{db}\g^{MA}\gd L,d,MB,\t^b & \gd L,a,BC,\t^C \\
\hline
\ell_{bd}\g^{AB}(2\gd H,d,CB,-\gd L,d,CB,)\t^C & \gd \wt\o,a,b,
\ema
\e
where
\bg3.44
\gd L,d,MB,=-\gd L,d,BM,\\
\ell_{dc}\g_{MB}\g^{CM}\gd L,d,CA, + \ell_{cd}\g_{AM}\g^{MC}\gd L,d,BC,
=2\ell_{cd}\g_{AM}\g^{MC}\gd H,d,BC,, \lb3.45
\e
$\gd L,d,CA,$ is Ad-type tensor \wrt $H$ (Ad-\ci\ on~$\ul P$)
\bg3.46
\gd \wt\o,a,b,=\gd \wt\G,a,bc,\t^c\\
\ell_{db}\gd \wt\G,d,ac,+\ell_{ad}\gd \wt\G,d,cb,=-\ell_{db}\gd C,d,ac, \lb3.47 \\
\gd \wt\G,d,ac,=-\gd \wt\G,d,ca,, \q \gd \wt\G,d,ad,=0. \lb3.48
\e
We define on $P$ a second \cn
\beq3.49
\gd W,\tA,\tB, = \gd \o,\tA,\tB, - \frac4{3(m+2)}\,\gd \d,\tA,\tB,\ov W.
\e
Thus we have on $P$ all $(m+4)$-\di al analogues of geometrical quantities from
NGT, i.e.
$$
\gd W,\tA,\tB,, \q \gd \o,\tA,\tB, \qh{and} \ka_{\tA\tB}.
$$

\section{GSW (Glashow--Salam--Weinberg) model in the \NK}\label{s:GSW}
Let $\ul P$ be a principal fiber bundle
\beq4.1
\ul P=(P,V,\pi,H,H)
\e
over the base space $V=E\tm S^2$ (where $E$ is a \spt, $S^2$---a two-\di al
sphere) with a projection~$\pi$, a \sc al group~$H$, a typical fiber~$H$
and a bundle manifold~$P$. We suppose that $H$ is semisimple. Let us define
on~$P$ a \cn~$\o$ which has values in a \Li\ of~$H,\fh$. Let us suppose that
a group $\SO(3)$ is acting on~$S^2$ in a natural way. We suppose that $\o$ is
\iv t \wrt an action of the group $\SO(3)$ on~$V$ in such a way that this
action is equivalent to $\SO(3)$ action on~$S^2$. This is equivalent to the
condition \er{3.23}. If we take a section $e:E\to P$ we get
\beq4.2
e^*\o=\gd A,a,A,\ov\t{}^A X_a=A_A\ov\t{}^A
\e
where $\ov\t{}^A$ is a frame on $V$ and $X_a$ are generators of the \Li~$\fh$.
\beq4.3
[X_a,X_b]=\gd C,c,ab,X_c.
\e
We define a \cvt\ of the \cn\ $\o$
\beq4.3a
\O=d\o+\frac12[\o,\o].
\e
Taking a section $e$
\bg4.4
e^*\O = \frac12 \,\gd F,a,AB,\ov\t{}^A \land \ov\t{}^B X_a =
\frac12 \,F_{AB}\ov\t{}^A \land \ov\t{}^B \\
\gd F,a,AB, = \pa_A\gd A,a,B, - \pa_B\gd A,a,A, - \gd C,a,cb, \gd A,c,A,
\gd A,c,B,. \lb4.5
\e

Let us consider a local \cd\ systems on~$V$. One has $x^A=(x^\mu,\psi,\vf)$
where $x^\mu$ are \cd\ system on~$E$, $\ov\t{}^\mu=dx^\mu$, and $\psi$
and~$\vf$ are polar and azimuthal angles on~$S^2$, $\t^5=d\psi$, $\t^6=d\vf$.
We have $A,B,C=1,2,\dots,6$, $\mu=1,2,3,4$. Let us introduce vector fields
on~$V$ corresponding to the infinitesimal action of $\SO(3)$ on~$V$
(see Ref.~\cite{21}). These
vector fields are called $\d_m=(\d^A_m)$, $\ov m=1,2,3$, $A=1,2,\dots,6$.
Moreover, they are acting only on the last two \di s ($A,B=5,6$, $\wt a,\wt
b=5,6$). We get:
\beq4.6
\bal
\d^\mu_{\bar m }&=0 &\hbox{and}&\\
\d^\psi_1&=\cos\vf, &&\d^\vf_1=-\cot\psi \sin\vf,\\
\d^\psi_2&=-\sin\vf, &&\d^\vf_2=-\cot\psi \cos\vf,\\
\d^\psi_3&=0, &&\d^\vf_3=1.
\eal
\e
They \sf y commutation relation of the \Li\ $A_1$ of a group $\SO(3)$,
\beq4.7
\d^A_{\bar m}\pa_A\d^B_{\bar n} - \d^A_{\bar n}\pa_A\d^B_{\bar m} =
\ve_{\bar m\bar n\bar p}\d^B_{\bar p}.
\e
The gauge field $A_A$ is spherically \s ic (\iv t \wrt an action of a group
$\SO(3)$) iff for some $V_{\bar m}$---a field on~$V$ with values in the
\Li~$\fh$---
\beq4.8
\pa_B\d^A_{\bar m}A_A + \d^A_{\bar m}\pa_A A_B = \pa_BV_{\bar m}
-[A_B,V_{\bar m}].
\e
It means that
\beq4.9
\mathop{\cL}_{\d_{\bar m}}A_A = \pa_BV_{\bar m} - [A_A,V_{\bar m}],
\e
a Lie \dv\ of $A_A$ \wrt $\d_{\bar m}$ results in a gauge \tf\ (see also
Eq.~\er{3.23}).

Eq.~\er{4.9} is \sf ied if
\beq4.10
V_1=\F_3\,\frac{\sin\vf}{\sin\psi}, \q
V_2=\F_3\,\frac{\cos\vf}{\sin\psi}, \q V_3=0
\e
and
\beq4.11
A_\mu=A_\mu(x), \q A_\psi=-\F_1(x)=A_5=\F_5, \q A_\vf = \F_2(x)\sin\psi
-\F_3\cos\psi=A_6=\F_6
\e
with the \fw\ constraints
\beq4.12
\bal
{}[\F_3,\F_1]&=-\F_2, \\
[\F_3,\F_2]&=\F_1, \\
[\F_3,A_\mu]&=0.
\eal
\e
$A_\mu,\F_1,\F_2$ are fields on $E$ with values in the \Li\ of~$H(\fh)$,
$\F_3$ is a \ct\ \el\ of Cartan subalgebra of~$\fh$. Let us introduce some
additional \el s according to the \E\nos\ Hermitian Kaluza--Klein Theory.
According to Section~\ref{s:els} we have on~$E$ a \nos\ Hermitian tensor $g_\m$, \cn s
$\gd \ov\o,\a,\b,$ and $\gd\ov W,\a,\b,$. On~$S^2$ we have a \nos\ metric
tensor
\beq4.13
\g_{\td a\td b}=r^2g_{\td a\td b}=r^2\bigl(\gd h,0,\td a\td b,
+\z \gd k,0,\td a\td b,\bigr)
\e
\setbox9=\hbox{$-\sin^2\psi$}%
\noindent where $r$ is the radius of a sphere $S^2$ and $\z$ is considered to
be pure imaginary,
\bea4.14
\gd h,0,\td a\td b,&=\left(\begin{array}{c|c}
\hbox to\wd9{\hfil $-1$\hfil} & 0 \\ \hline 0 & -\sin^2\psi \end{array}\right)\\
\gd k,0,\td a\td b,&=\left(\begin{array}{c|c}
0 & \hbox to\wd9{\hfil$\sin\psi$\hfil} \\ \hline -\sin\psi & 0 \end{array}\right) \lb4.15
\e
\setbox0=\hbox{$-\z\sin\psi$}%
\def\vp{\vrule height\ht0 depth\dp0 width0pt}%
\noindent and a \cn\ compatible with this \nos\ metric
\bg4.16
g_{\td a\td b}=
\begin{array}{c}
\noalign{\vskip-10pt}
\begin{array}{ccc}
\hbox to\wd0{\hfil$\scriptstyle 5$\hfil}&\hbox to\wd0{\hfil$\scriptstyle 6$\hfil}\end{array}\\
\begin{array}{cc}
\left(\begin{array}{c|c}
-1 & \z\sin\psi \\ \hline -\z\sin\psi & -\sin^2\psi
\end{array}\right)&
\begin{array}{c}
\scriptstyle 5\vp\\\scriptstyle 6\vp
\end{array}
\end{array}
\end{array}\\
\wt g=\det(g_{\td a\td b}) = \sin^2\psi(1+\z^2) \lb4.17 \\
g^{\td a\td b} = \frac1{\sin^2\psi(1+\z^2)}
\begin{array}{c}
\noalign{\vskip-4pt}
\begin{array}{ccc}
\hbox to\wd0{\hfil$\scriptstyle 5$\hfil}&\hbox to\wd0{\hfil$\scriptstyle 6$\hfil}\end{array}\\
\begin{array}{cc}
\left(\begin{array}{c|c}
-\sin^2\psi & -\z\sin\psi \\ \hline \z\sin\psi & -1
\end{array}\right)&
\begin{array}{c}
\scriptstyle 5\vp\\\scriptstyle 6\vp
\end{array}
\end{array}
\end{array}, \lb4.18
\e
$\wt a,\wt b=5,6$. In this way we have to do with K\"ahlerian structure
on~$S^2$ (Riemannian, symplectic and complex which are compatible). This seems to
be very interesting in further research connecting \un\ of all fundamental
\ia s. On $H$ we define a \nos\ metric
\beq4.19
\ell_{ab}=h_{ab}+\xi k_{ab}
\e
where $k_{ab}$ is a right-\iv t skew-\s ic 2-form on~$H$.

One can rewrite the constraints \er{4.12} in the form
\beq4.20
\bal
{}[\F_3,\F]&=i\F\\
[\F_3,\wt\F]&=-i\wt\F\\
[\F_3,A_\mu]&=0
\eal
\e
where $\F=\F_1+i\F_2$, $\wt\F=\F_1-i\F_2$ (see Ref.~\cite{21}).

In this way our 6-\di al gauge field (a \cn\ on a fiber bundle) has been
reduced to a 4-\di al gauge one (a~\cn\ on a fiber bundle over a \spt~$E$)
and a collection of scalar fields defined on~$E$ \sf ying some constraints.
According to our approach there is defined on~$S^2$ a \nos\ \cn\ compatible
with a \nos\ tensor $g_{\td a\td b}$, $\wt a,\wt b=5,6$,
\beq4.21
\bga
\wh Dg_{\td a\td b}=g_{\td a\td d}\gd Q,\td d,\td b\td c,(\wh\G)\ov\t{}
^{\td c}\\
\gd Q,\td d,\td b\td d,(\wt\G)=0
\ega
\e
where $\wh D$ is an exterior \ci\ \dv\ \wrt a \cn\ $\gd\wh\o,\td a,\td b,
=\gd\wh\G,\td a,\td b\td c,\ov\t{}^{\td c}$ and $\gd Q,\td d,\td b\td c,
(\wh \G)$ its torsion.

Let us metrize a bundle $P$ in a \nos\ way. On~$V$ we have \nos\ tensor (see
Ref.~\cite1)
\beq4.22
\g_{AB}=\bma g_\m & 0 \\\hline 0 & r^2g_{\td a\td b} \ema
\e
and a \nos\ \cn\ $\gd\ov\o,A,B,=\gd \G,A,BC,\t^C$ compatible with this tensor
\beq4.23
\bga
\ov D\g_{AB}=\g_{AD}\gd Q,D,BC,(\ov\G)\t^C\\
\gd Q,D,BD,(\ov\G)=0.
\ega
\e
The form of this \cn\ is as follows
\beq4.24
\gd \ov\o,A,B,=\bma \gd \ov\o,\a,\b, & 0 \\\hline 0 & \gd \wh\o,\td a,\td b,
\ema
\e
where $\ov D$ is an exterior \ci\ \dv\ \wrt $\gd \ov\o,A,B,$ and $\gd Q,D,BC,
(\ov\G)$ its torsion.

Afterwards we define on $P$ a \nos\ tensor
\beq4.25
\ka_{\tA\tB}\t^\tA \ot \t^\tB = \pi^* (\g_{AB}\ov\t{}^A \ot \t^B)
+\ell_{ab}\t^a \ot \t^b
\e
where
\beq4.26
\t^\tA = (\pi^*(\ov \t{}^A),\la\o^a),
\e
$\o=\o^0X_a$ is a \cn\ defined on $P$ ($\tA,\tB,\tC=1,2,\dots,n+6$).

We define on $P$ two \cn s $\gd \o,A,B,$ and $\gd W,A,B,$ \st $\gd \o,A,B,$
is compatible with a \nos\ tensor $\ka_{\tA\tB}$,
\beq4.27
\bga
D\ka_{\tA\tB}=\ka_{\tA\td D}\gd Q,\td D,\tB\tC,(\G)\t^\tC\\
\gd Q,\td D,\tB\td D,(\G)=0,
\ega
\e
where $D$ is an exterior \ci\ \dv\ \wrt a \cn\ $\gd \o,\tA,\tB,$ and
$\gd Q,\td D,\tB\tC,(\G)$ its torsion.

The second \cn
\beq4.28
\gd W,\tA,\tB,=\gd \o,\tA,\tB, - \frac4{3(n+4)}\,\gd \d,\tA,\tB,\ov W
\q (n=\dim H).
\e

In this way we have all quantities known from Section~\ref{s:els}. We calculate a
scalar of \cvt\ (\MR) for a \cn\ $\gd W,\tA,\tB,$ and afterwards an action
\bml4.29
S=-\frac1{V_1V_2} \int_U \sqrt{-g}\,d^4x \int_H\sqrt{|\ell|}\,d^nx
\int_{S^2}\sqrt{|\wt g|}\,d\O\,R(W) \\
{}=-\frac1{r^2V_1V_2} \int_U \sqrt{-g}\,d^4x \int_{S^2}\sqrt{|\wt g|}\,d\O\,
\Bigl(\ov R(\ov W)\\
{}+\frac{8\pi G_N}{c^4}\Bigl(\cLY+\frac1{4\pi r^2}\,
\cL_{\rm kin}(\n\F)-\frac1{8\pi r^2}\,V(\F)-\frac1{2\pi r^2}\,
\cL_{\rm int}(\F,\wt A)\Bigr)+\la_c\Bigr)
\e
where $V_1=\int_U\sqrt{|\ell|}\,d^nx$, $V_2=\int_{S^2}\sqrt{|\wt g|}\,d\O$,
$U\subset E$,
\beq4.30
\la_c=\Bigl(\frac{\a_s^2}{\ell\pl^2}\,\wt R(\wt \G)+\frac1{r^2}\,\ul{\wt P}
\Bigr)
\e
where $\wt R(\wt\G)$ is a \MR\ \cvt\ scalar on a group~$H$ (see Section~3 for
details).
\beq4.32
\wt {\ul P}=\frac1{V_2}\int_{S^2}\sqrt{|\wt g|}\,d\O\,\wh R(\wh \G)
\e
where $\wh R(\wh \G)$ is a \MR\ \cvt\ scalar on $S^2$ for a \cn\
$\gd\wh\o,\td a,\td b,$.
\beq4.33
\cLY=-\frac1{8\pi}\,\ell_{ij}\bigl(\wt H{}^{(i}\wt H{}^{j)}- \wt L^{i\m}
\gd\wt H,j,\m,\bigr)
\e
where
\beq4.34
\ell_{ij}g_{\mu\b}g^{\g\mu}\gd\wt L,i,\g\a, + \ell_{ji}g_{\a\mu}g^{\mu\g}
\gd \wt L,i,\b\g, = 2\ell_{ji}g_{\a\mu}g^{\mu\g}\gd\wt H,i,\b\g,
\e
One gets from \er{3.44}
\bg4.39
\gd L,b,\td a\td b,=h^{bc}\ell_{cd}\gd H,d,\td b\td a,,\\
V(\F)=-\frac1{V_2}\int\sqrt{|\wt g|}\,d\O\,\bigl(2h_{cd}(\gd H,c,\td a\td b,
g^{\td a\td b})(\gd H,d,\td c\td d,g^{\td c\td d}) - \ell_{cd}g^{\td a\td m}
g^{\td b\td n}\gd L,c,\td a\td b,\gd H,d,\td m\td n,\bigr)\hskip90pt \nn\\
\hskip90pt{}=\frac1{V_2}\,\frac{2\pi^2}{\sqrt{1+\z^2}}\,\ka\bigl(
(\ve_{\bar r\bar s\bar t}\F_{\bar t}+[\F_{\bar r},\F_{\bar s}]),
(\ve_{\bar r\bar s\bar t}\F_{\bar t}+[\F_{\bar r},\F_{\bar s}])\bigr)
\lb4.35 \\
\ka_{de}=(1-2\z^2)h_{de}+\xi^2\gd k,c,d,k_{ce} \lb4.36
\e
where
\bg4.37
\gd k,c,d,=h^{cf}k_{fd} \\
V_2=\int_{S^2}\sqrt{|\wt g|}\,d\O=4\pi \sqrt{1+\z^2}, \lb4.38
\e
$\ov r,\ov s,\ov t=1,2,3$, $\ve_{\bar r\bar s\bar t}$ is a usual anti\s ic
symbol $\ve_{123}=1$.

We get also from \er{3.44}
\beq4.40
\ell_{dc}g_{\mu\b}g^{\g\mu}\gd L,d,\g\td a, + \ell_{cd}\gd L,d,\b\td a,
=2\ell_{cd}\gd F,c,\b\td a,.
\e
Using the \e\ (see Ref.~\cite5)
\bmlg
\gd L,n,\o\td m,=\gv{\n_\o}\F^n_{\td m}+ \xi \gd k,n,d,\gv{\n_\o}
\F^d_{\td m} - \bigl(\z\gv{\n_\o}\F^n_{\td a}h^{0\td a\td d}k_{0\td d\td m}
+\wt g{}^\(\a\mu)\gv{\n_\a}\F^n_{\td m}g_\[\mu\o]\bigr)\\
{}-2\xi\z\gd k,n,d,\gv{\n_\o}\F^d_{\td d}\wt g{}^\(\d\a)g_\[\a\o]
h^{0\td d\td a}\gd k,0,\td a\td m,+\xi\gd k,n,d,\bigl(\z^2h^{\td d\td a}
\gv{\n_\o}\F^d_{\td a}\gd k,0,\td d\td b,\gd k,0,\td m\td c,h^{0\td c\td b}\\
{}+\gv{\n_\b}\F^d_{\td m}\wt g{}^\(\d\b)g_\[\d\a]g_\[\o\mu]\wt g{}^\(\a\mu)
\bigr)
-\xi^2k^{nb}k_{bd} \bigl(\z\gv{\n_\o}\F^d_{\td a}h^{0\td a\td b}\gd k,0,\td
m\td b, + \wt g{}^\(\a\b)\gv{\n_a} \F^d_{\td m}g_\[\o\b]\bigr)
\e
where
$$
k^{nb}=h^{na}h^{bp}k_{ap},
$$
one gets
\bml4.41
\gd L,n,\o\td m,=\gv{\n_\o}\Ft nm + \xi\gd k,n,d, \gv{\n_\o}\Ft dm
-\wt g{}^\(\a\mu)\gv{\n_\a}\Ft nm g_\[\mu\o]\\
{}+\xi \gd k,n,d,\gv{\n_\b}\Ft dm\wt g{}^\(\d\b)g_\[\d\a]g_\[\o\mu]
\wt g{}^\(\a\mu) - \xi^2 k^{nb}k_{bd}\wt g{}^\(\a\b)\gv\n \Ft dmg_\[\o \b].
\e
Moreover, now we have to do with Minkowski space $g_\m=\eta_\m$ and
\beq4.42
\gd L,n,\o\td m,=\gd H,n,\o\td m,+\xi \gd k,n,d,\gd H,d,\o\td m,.
\e
We remember that $\wt m=5,6$ or $\vf,\psi$ and that
\beq4.43
\gd H,n,\mu\td m,=\gv{\n_\mu}\Ft nm.
\e
We have
\beq4.44
\cL_{\rm kin}(\gd H,n,\mu\td m,)=\frac1{V_2} \int \sqrt{|\wt g|}\,d\O\,
(\ell_{ab}\eta^{\b\mu}\gd L,a,\b\td b,\gd H,b,\mu\td a,g^{\td b\td a}).
\e
Finally we get
\bg4.45
\cL_{\rm kin}(\n_\mu\F_{\bar m})=\frac{2\pi^2}{V_2}\,\frac{\eta^\m}
{\sqrt{1+\z^2}}\,\bar\ka\bigl(\gv{\n_\mu}\F_{\bar m},\gv{\n_\nu}\F_{\bar m}
\bigr)\\
\ov\ka_{ad}=(h_{ad}+\xi^2k_{ab}\gd k,b,d,) \lb4.46
\e
where
\beq4.47
\gv{\n_\mu}\F_{\bar m}=\pa_\mu\gd \F,a,\bar m, - [A_\mu,\F_{\bar m}].
\e

Now we follow Ref.~\cite{21} and suppose $\mathop{\rm rank}H=2$ and
afterwards $H=G2$. In this way our lagrangian can go to the GSW model where
$\SU(2) \tm \U(1)$ is a little group of~$\F_3$ (see Appendix~B).
We get also a Higgs' field
complex doublet and \sn\ \s y breaking and mass generation for intermediate
bosons. For simplicity we take $\xi=0$ and also we do not consider an
influence of the \nos\ gravity on a Higgs' field.
We get also a mixing angle $\t_W$ (Weinberg angle). If we choose
$H=G2$ we get $\t_W=30^\circ$. We get also some predictions of masses
\beq4.49
\frac{M_H}{M_W} = \frac1{\cos\t_W}\cdot \sqrt{1-2\z^2}
\e
where $\z$ is an arbitrary \ct
\beq4.50
\frac{M_H}{M_W} = \frac{2\sqrt{1-2\z^2}}{\sqrt3}.
\e
We take $M_H\simeq125$\,GeV and $M_W\simeq80$\,GeV (see Refs
\cite{26,27,28,29,30}).

One gets
\beq4.51
\z=\pm 0.911622i.
\e
Thus $\z$ is pure imaginary.
This means we can explain mass pattern in GSW model. $r$ gives us a scale of
mass and is an arbitrary parameter.

Moreover, a scale of energy is equal to
$M=\frac{\hbar c}{r\sqrt{2\pi}\,\sqrt{1+\z^2}}$
which we equal to MEW (electro-weak) energy scale, i.e.\ to~$M_W$. One gets
$r\simeq 2.39\tm 10^{-18}$\,m. In the original Manton model Higgs' boson is too
light. We predict here masses for $W,Z^0$ and Higgs bosons in the theory taking
two parameters, $\z$ (Eq.~\er{4.51}) and $r\simeq 2.39\tm 10^{-18}$\,m in order
to get desired pattern of masses. The value of the Weinberg angle derived here
for $H=G2$ has nothing to do with ``GUT driven'' value $\frac14$ for $\frac14$
is a value of our $\sin^2\t_W$, not $\sin\t_W$.
According to Ref.~\cite{21} a Lie group $H$ should have a Lie algebra $\fh$
with rank~2. We have only three possibilities: G2, $\SU(3)$ and $\SO(5)$.
The angle between two roots plays a role of a Weinberg angle. For $\SO(5)$
$\t=45^\circ$ and for $\SU(3)$ $\t=60^\circ$. Only for $G2$, $\t=\t_W
=30^\circ$, which is close to the experimental value. In this way a \un\
chooses $H=G2$.

Let us notice that $\dim G2=14$ and for this $\dim P=20$.

Moreover, we have
\beq4.52
M_Z = \frac{M_W}{\cos\t} = \frac{M_W}{\cos\t_W} = \frac2{\sqrt3}\,M_W
\simeq 92.4
\e
and we get from the theory
\beq4.53
\sin^2\t_W=0.25 \q (\t_W=30^\circ).
\e
However from the experiment we  get
\beq4.54
\sin^2\t_W=0.2397\pm0.0013
\e
which is not $0.25$.

Moreover, from theoretical point of view the value $0.25$ is a value without
radiation corrections and it is possible to tune it at $Q=91.2$\,GeV/c in the
$\ov{\rm MS}$ scheme to get the desired value.

Let us notice the following fact. In the electroweak theory we have a
Lagrangian for neutral current \ia
\bml4.55
\cL_N = q J^{\rm em}_\mu A^\mu + \frac g{\cos\t_W}(J^3_\mu - \sin^2\t_W
J^{\rm em}_\mu)Z^{0\mu}
=qJ^{\rm em}_\mu A^\mu +\sum_f \ov\psi_f \g_\mu(g^f_V-g^f_A\g^5)\psi_f
Z^{0\mu}
\e
where $g^f_V$ and $g^f_A$ are coupling \ct s for vector and axial \ia s for a
fermion~$f$. One gets
\beq4.56
\bal
g^f_V&=\frac{2q}{\sin2\t_W}(T^3_f - 2q_f\sin^2\t_W)\\
g^f_A&=\frac{2q}{\sin2\t_W}
\eal
\e
where $T^3_f$ is the third component of a weak isospin of a fermion~$f$ and
$q_f$ is its electric charge measured in \el ary charge~$q$,
\beq4.57
q_f=T^3_f+\frac{Y_f}2
\e
where $Y_f$ is a weak hypercharge for~$f$. It is easy to see that for an
electron we get $g^f_V=0$ if $\t_W=30^\circ$.

Moreover, we know from the experiment that
\beq4.58
g^f_V\ne0
\e
(see Ref.\ \cite{26}).

Following Ref.~\cite{21} we use the following formulae
\bea B.1
\F_5&=\frac12(\vf^*_1x_{-\a}+\vf^*_2x_{-\b}-\vf_1x_\a-\vf_2x_\b)\\
\F_6&=\frac{\sin\psi}{2i} (\vf_1x_\a+\vf_2x_\b+\vf^*_1x_{-\a}+\vf^*_2x_{-\b})
-\F_3\cos\psi. \lb B.2
\e
$\F_3$ is \ct\ and commutes with a reduced \cn. $\SU(2)\tm U(1)$ is a little
group of~$\F_3$,
\beq B.3
\F_3=\frac12\,i(2-\langle \g,\a\rangle)^{-1}(h_\a+h_\b),
\e
$x_\a$, $x_{-\a}$, $x_\b$, $x_{-\b}$ are \el s of a \Li~$\fh$ of~$H$ (see Ref.~\cite{b})
corresponding to roots $\a,-\a,\b,-\b$, $h_\a$~and~$h_\b$ are \el s of Cartan
subalgebra of~$\fh$ \st
\beq B.4
h_\a = \frac{2\a_i}{\a\cdot\a}\,H_i = [x_\a,x_{-\a}],
\e
where $\a=(\a_1,\dots,\a_k)$, $k=\mathop{\rm rank}(\fh)$, $\g=\a-\b$,
$[H_i,x_\o]=\o_ix_\o$, $H_i$~form Cartan subalgebra of~$\fh$, $[x_\o,x_\tau]
=C_{\o,\tau}x_{\o+\tau}$ if $\o+\tau$ is a root, if $\o+\tau$ is not a root
$x_\o$ and $x_\tau$ commute. We take $k=2$.
\beq B.5
\langle \g,\a \rangle = \frac{2\g\cdot\a}{\a\cdot \a} =
2\,\frac{|\g|}{|\a|}\cos\t.
\e
In this way we get a Higgs' doublet $\binom{\vf_1}{\vf_2}=\wt\vf$.

\allowdisplaybreaks
The $\SU(2)\tm \U(1)$ generators are given by
\beq B.6
\bal
t_1&=\frac12\,i(x_\g+x_{-\g})\\
t_2&=\frac12(x_\g-x_{-\g})\\
t_3&=\frac12\,ih_\g\\
y&=\frac12\,ih.
\eal
\e
$h$ is an \el\ of Cartan subalgebra orthogonal to $h_\g$ with the same norm.
Now everything is exactly the same as in Ref.~\cite{21} except the fact that
\bea B.7
\bar k_{ad}&=h_{ad}-\xi^2 k_{ab}\gd k,b,d,\\
k_{ad}&=(1-2\z^2)h_{ad}-\xi^2k_{ab}\gd k,b,d,. \lb B.8
\e
In Ref.~\cite{21}
\beq B.9
\bar k_{ad}=k_{ad}=h_{ad}.
\e
A four-\pt\ of \YM' field (a \cn\ $\o_E$) can be written as
\bg B.10
A_\mu=\sum_{i=1}^3 A_\mu t_i + B_\mu y \\
\hbox{or}\q
A_\mu=\frac12 \, i(A_\mu^- x_\g + A_\mu^+ x_{-\g} + A_\mu^3 h_\g
+ B_\mu h) \lb B.11 \\
A_\mu^\pm = A_\mu^1 \pm i A_\mu^2. \lb B.12
\e
We have (see Ref.~\cite{21})
\begin{align}
{}&h(t_i,t_j)=-\frac1{\g\cdot\g}\,\d_{ij} \nn\\
&h(y,y)=-\frac1{\g\cdot\g} \nn\\
&h(t_i,y)=0\nn\\
&F_\m=\bigl(\pa_\mu A_\nu^a - \pa_\nu A^a_\mu + \gd\ve,a,bc,A_\mu^b
A_\nu^c\bigr)t_a + (\pa_\mu B_\nu-\pa_\nu B_\mu )y
= \gd F,a,\m,t_a + B_\m y \lb B.13 \\
&h(F_\m,F_\m)=-\frac{\d_{ab}}{\g\cdot\g}\,\gd F,a,\m,F^{b\m}-
\frac1{\g\cdot\g}B_\m B^\m \lb B.14 \\
&\gv{\n_\mu}\F=\Bigl(\pa_\mu \vf_1 - \frac12 \,iA_\mu^- \vf_2
-\frac12\,i A_\mu^3 \vf_1 
-\frac12\,i\tan \t B_\mu \vf_1\Bigr)x_\a \nn \\
&\hskip30pt {}+\Bigl(\pa_\mu \vf_2 - \frac12 \,iA_\mu^+ \vf_1
+\frac12\,iA_\mu^3\vf_2 -\frac12\,i\tan \t B_\mu
\vf_2\Bigr)x_\b \lb B.15 \\
&\gv{\n_\mu}\wt{\F} = -\Bigl(\pa_\mu \vf_1^* + \frac12 \,iA_\mu^+ \vf_2^*
+\frac12\,i A_\mu^3 \vf_1^* +\frac12\,i\tan \t B_\mu \vf_1^*\Bigr)x_{-\a}\nn\\
&\hskip30pt {}-\Bigl(\pa_\mu \vf_2^* + \frac12 \,iA_\mu^- \vf_1^*
-\frac12\,i A_\mu^3 \vf_2^* +\frac12\,i\tan \t B_\mu\vf_2^*\Bigr)x_{-\b}
\lb B.16
\end{align}
We redefine the fields $A_\mu^a$, $B_\mu$ and $\wt\vf$ with some rescaling
($g$ is a coupling \ct)
\beq B.17
A_\mu^{\prime a}  = L_1 A_\mu^a, \q B_\mu'  = L_1 B_\mu, \q
\wt\vf{}'  = L_2\wt\vf
\e
where
\bea B.18
L_1 &= \frac1g\,\frac1{(\g\cdot \g)^{1/2}}\\
L_2 & = \frac1g\Bigl(\frac{\g\cdot\g}{\a\cdot\a}\Bigr)^{1/2} \lb B.19
\e
We proceed the following \tf
\beq B.20
\left(\begin{matrix}
Z_\mu^0 \\ A_\mu \end{matrix}\right) =
\left(\begin{matrix}
\cos\t &\ & -\sin\t \\ \sin\t && \cos \t \end{matrix}\right)
\left(\begin{matrix}
A_\mu^3 \\ B_\mu \end{matrix}\right).
\e
\goodbreak

According to the classical results we also have $\frac {g'}g = \tan\t$,
assuming $q=g\sin\t$, where $q$ is an \el ary charge and $g$~and~$g'$ are
coupling \ct s of $A_\mu^a$ and~$B_\mu$ fields. The \sn\ \s y breaking and
\Hm\ in the Manton model works classical if we take for minimum of the \pt
\beq B.21
\wt\vf_0=\left(\begin{matrix} 0 \\ \frac v{\sqrt2} \end{matrix}\right)
e^{i\a}, \q \a \hbox{ arbitrary phase,}
\e
and we parametrize $\wt\vf=\binom{\vf_1}{\vf_2}$ in the following way
\beq B.22
\wt\vf(x)=\exp\Bigl(i\,\frac1{2v}\,\si^a t^a(x)\Bigr)
\left(\begin{matrix} 0 \\ \frac {v+H(x)}{\sqrt2} \end{matrix}\right).
\e
For a vacuum state we take
\beq B.23
\wt\vf_0=\left(\begin{matrix} 0 \\ \frac {v}{\sqrt2} \end{matrix}\right),
\e
$t^a(x)$ and $H(x)$ are real fields on~$E$. $t^a(x)$ has been ``eaten'' by
$A_\mu^a$, $a=1,2$, and~$Z_\mu^0$ fields making them massive. $H(x)$~is our
Higgs' field. $\si^a$~are Pauli matrices.

In the formulae \er{B.7}--\er{B.8} we take $\xi=0$. One gets in the
Lagrangian mass terms:
$$
M_W^2 W_\mu^+ W^{-\mu} + \frac12\, M_Z^2 Z_\mu^0Z^{0\mu} - \frac12\,M_H^2H^2,
$$
where $W_\mu^+=A_\mu^+$, $W_\mu^-=A_\mu^-$, getting masses for
$W^\pm$, $Z^0$ bosons and
a~Higgs boson (see Eqs \er{4.49}--\er{4.53}). For G2 $\langle \g,\a
\rangle=3$ and $\t=30^\circ$, $\t$~is identified with the Weinberg angle
$\t_W$.

In order to proceed a \Hm\ and \sn\ \s y breaking in this model we use the
following gauge \tf
\beq B.23a
\wt\vf (x) \mapsto U(x)\wt\vf(x)=\frac1{\sqrt2}
\left(\begin{matrix} 0 \\ v+H(x) \end{matrix}\right),
\e
where
\beq B.24
v=\frac{2\sqrt2}{rg}\cos\t
\e
a vacuum value of a Higgs field
\beq B.25
U(x)=\exp\Bigl(-\frac1{2v}\,t^a(x)\si^a\Bigr).
\e
$H(x)$ is the remaining scalar field after a \s y breaking and a \Hm. One gets
\bg B.26
A_\mu \mapsto A_\mu^u = {\rm ad}'_{U^{-1}(x)}A_\mu + U^{-1}(x)\pa_\mu U(x)\\
F_\m \mapsto F_\m^u = {\rm ad}'_{U^{-1}(x)}F_\m. \lb B.27
\e

Using some
additional fields $\F_1,\F_2,\F_3$ and also $\F$ and~$\wt\F$, we can write
$\gv{\n_\mu}\F_5$ and $\gv{\n_\mu}\F_6$ in terms of Higgs' fields $\vf_1$
and~$\vf_2$,
\begin{align}
\gv{\n_\mu}\F_5&=\frac12\gv{\n_\mu}(\F+\wt\F)=\frac12\Bigl[\Bigl(\pa_\mu\vf_1
-\frac12\,iA_\mu^- \vf_2 - \frac12\,iA_\mu^3 \vf_1 -\frac12\,i\tan\t B_\mu
\vf_1\Bigr)x_\a \nn \\
&+\Bigl(\pa_\mu \vf_2 - \frac12\,iA_\mu^+\vf_1 + \frac12\,iA_\mu^+\vf_2
-\frac12\,iB_\mu\vf_2\tan\t\Bigr)x_\b \nn \\
&-\Bigl(\pa_\mu\vf_1^* + \frac12\,iA_\mu^+ \vf_2^* + \frac12\,iA_\mu^3\vf_1^*
+\frac12\,iB_\mu\vf_1^*\tan\t\Bigr)x_{-\a} \nn \\
&-\Bigl(\pa_\mu\vf_2^* + \frac12 \,iA_\mu^- \vf_1^*
-\frac12\,i A_\mu^3 \vf_2^* +\frac12\,i\tan \t B_\mu\vf_2^*\Bigr)x_{-\b}\Bigr]
\lb C.4a
\end{align}
\begin{align}
\gv{\n_\mu}\F_6 &= \frac{\sin\psi}{2i} \gv{\n_\mu}(\F-\wt\F)=
\frac{\sin\psi}{2i}
\Bigl[\Bigl(\pa_\mu\vf_1
-\frac12\,iA_\mu^- \vf_2 - \frac12\,iA_\mu^3 \vf_1 -\frac12\,i\tan\t B_\mu
\vf_1\Bigr)x_\a \nn \\
&+\Bigl(\pa_\mu \vf_2 - \frac12\,iA_\mu^+\vf_1 + \frac12\,iA_\mu^+\vf_2
-\frac12\,iB_\mu\vf_2\tan\t\Bigr)x_\b \nn \\
&-\Bigl(\pa_\mu\vf_1^* + \frac12\,iA_\mu^+ \vf_2^* + \frac12\,iA_\mu^3\vf_1^*
+\frac12\,iB_\mu\vf_1^*\tan\t\Bigr)x_{-\a} \nn \\
&-\Bigl(\pa_\mu\vf_2^* + \frac12 \,iA_\mu^- \vf_1^*
-\frac12\,i A_\mu^3 \vf_2^* +\frac12\,i\tan \t B_\mu\vf_2^*\Bigr)x_{-\b}\Bigr]
\lb C.5
\end{align}
where
\bg C.18
e^*\o_E = A^i_\mu \ov \t{}^\mu t_i + B_\mu \ov \t{}^\mu y \\
e^*\o = \a^c_i A^i_\mu \ov \t{}^\mu \wt t_i + \F^a_{\td a}\t^{\td a}X_a ,
\lb C.20 \\
\wt t_i=t_i,\ i=1,2,3, \q \wt t_4=y. \lb C.21
\e

Let us proceed a \sn\ \s y breaking and \Hm.
In this way we transform
\beq C.26
\gv{\n_\mu}\F_{\td a} \mapsto {\rm ad}'_{U^{-1}(x)}\gv{\n_\mu}\F_{\td a} = \gv{\n_\mu}\F_{\td a}^u\,,
\q \wt a=5,6,
\e
where
\begin{gather}
\gv{\n_\mu}\F_5^u = \frac1{2\sqrt2}\Bigl[\pa_\mu H(x)(x_\b-x_{-\b})\hskip200pt
\nn\\
\hskip40pt{}+\frac i2(v+H(x))\bigl(A_\mu^{3u}(x_\b+x_{-\b})+B_\mu
\tan\t(x_{-\b}-x_\b)
-A_\mu^{+u}x_{-\a}+A_\mu^{-u}x_\a\bigr)\Bigr] \lb C.27 \\
\gv{\n_\mu}\F_6^u = \frac{\sin\psi}{2i}\Bigl[\pa_\mu H(x)(x_\b+x_{-\b})
\hskip200pt \nn \\
\hskip40pt{}+\frac i2(v+H(x))\bigl(A_\mu^{+u}x_{-\a}-A_\mu^{-u}x_\a
+A_\mu^{3u}(x_\b-x_{-\b})+B_\mu \tan\t(x_{-\b}-x_\b)\bigr)\Bigr] \lb C.28
\end{gather}
where
\bg C.30
H_{56}^u =-\frac{\sin\psi(v+H(x))}2 \Bigl((v+H(x))\frac{\b_i}{\b\cdot \b}
\,H_i +\sqrt2 \cos\psi(x_\b+x_{-\b})\Bigr) \\
H_{56}^u = - H_{65}^u \lb C.31
\e
where
\bea C.32
A_\mu^+ \mapsto A_\mu^{+u} &= \bigl({\rm ad}'_{U^{-1}(x)}A_\mu\bigr)^+
+\frac i{2v}\,\pa_\mu t^+(x) \\
A_\mu^- \mapsto A_\mu^{-u} &= \bigl({\rm ad}'_{U^{-1}(x)}A_\mu\bigr)^-
+\frac i{2v}\,\pa_\mu t^-(x) \lb C.33 \\
A_\mu^3 \mapsto A_\mu^{3u} &= \bigl({\rm ad}'_{U^{-1}(x)}A_\mu\bigr)^3
+\frac i{2v}\,\pa_\mu t^3(x). \lb C.33b
\e

Let us suppose that $H=G2$. In this case one gets
\beq C.52
\bga
|\b|=|\a|=\sqrt2, \q |\g|=\sqrt6,\\
\a\cdot\a=\b\cdot\b=2, \q \g\cdot\g=6,\\
\langle \g,\a \rangle=3, \q \langle \g,\b \rangle=\langle \a,\b \rangle=-1,\\
\frac{\g_1\a_2-\g_2\a_1}{\g\cdot\g}=\frac{\sqrt3}6,\\
\t=30^\circ, \q \cos\t=\frac{\sqrt3}2\,, \q \sin\t=\frac12\,.
\ega
\e

\section{Spinor fields on $P$}\label{s:spinor}
Let $\Psi$ be a spinor field on $P$ belonging to \fn\ \rp ation $D^F$ of $\SO(1,n+3)$
($\Spin(1,n+3)$) or $\SO(1,n+n_1+3)$ ($\Spin(1,n+n_1+3)$) or also $\SO(1,19)$ ($\Spin(1,19)$)
and $\G^A$, $A=1,2,\dots,n+4$ be a \rp ation of the Clifford algebra for $\SO(1,n+3)$ acting
in the space \rp ation $D^F$ (see Refs \cite{ax1}, \cite{ax2}, \cite{ax3}), i.e., $\G^A\in C(1,n+3)$,
\beq 14.1
\bga
\{\G^A,\G^B\}=2\eta_{AB},\\
\G^A\in L(\C^k), \q k=4\cdot 2^{[n/2]}, \q [n/2]=l,\\
\eta_{AB}=\diag(-1,-1,-1,1,\underbrace{-1,-1,\dots,-1}_{n\ \rm times}).
\ega
\e
We consider also two additional cases mentioned above:
$$
\{\G^\tA,\G^\tB\} = 2\eta_{\tA\tB}, \q \G^\tA \in L(\C^k),
$$
where
$$
k=4\cdot2^{[(n+n_1)/2]}, \q \eta_{\tA\tB}=(-1,-1,-1,1,\underbrace{-1,-1,\dots,-1}_{n+n_1\ \rm times}).
$$
We introduce a spinor field $\ov\Psi$
\beq14.2
\ov\Psi = \Psi^+ B
\e
where ``$+$'' is a Hermitian conjugation and
\beq14.3
\G^{A+} = B\G^A B^{-1}.
\e
Usually $B=\G^4$.

It is easy to see that
\beq14.4
\ov\Psi(pg_1) = \ov\Psi(p)\si(g_1),
\e
$p=(x,g)\in\ul P$, $g,g_1\in G$ (or $H$), $\si$ is a unitary \rp ation of the group~$G$ (or~$H$) acting in $k=4\cdot 2^l$-\di al
complex space $\si\in L(\C^k)$. The fields $\Psi$ and~$\ov\Psi$ are defined on~$\ul P$ and $P$~is assumed to have an orthonormal \cd\
system $\t^A$ (or~$\t^\tA$). This \cd\ system is in general non-holonomic. We perform an infinitesimal change of the frame $\t^A$
(or~$\t^\tA$)
\beq14.5
\t^{A'} = \t^A + \d\t^A = \t^A - \gd\ve,A,B, \t^B, \q \ve_{AB}+\ve_{BA}=0
\e
or
\beq14.6
\t^{\tA'} = \t^\tA + \d\t^\tA = \t^\tA - \gd\ve,\tA,\tB, \t^\tB, \q \ve_{\tA\tB}+\ve_{\tB\tA}=0.
\e
Suppose that the field $\Psi$ corresponds to $\t^A$ (or $\t^\tA$) and $\Psi'$ to $\t^{A'}$ ($\t^{\tA'}$), then we get
\beq14.7
\bga
\Psi' = \Psi + \d\Psi = \Psi - \ve_{AB}\wh\si{}^{AB}\Psi,\\
\ov\Psi' = \ov\Psi + \d\ov\Psi = \ov\Psi + \ov\Psi\ve_{AB}\wh\si{}^{AB},
\ega
\e
where
$$
\wh\si{}^{AB} = \tfrac18 [\G^A,\G^B].
$$
Simultaneously we have
\beq14.8
\bga
\Psi' = \Psi + \d\Psi = \Psi - \ve_{\tA\tB}\wh\si{}^{\tA\tB}\Psi,\\
\ov\Psi' = \ov\Psi + \d\ov\Psi = \ov\Psi + \ov\Psi\ve_{\tA\tB}\wh\si{}^{\tA\tB},\\
\wh\si^{\tA\tB} = \tfrac18 [\G^{\tA},\G^{\tB}].
\ega
\e

Now we consider \ci\ \dv s of spinor fields $\Psi$ and $\ov\Psi$ on~$P$ \wrt a \cn\ $\gd\wt\o,A,B,$ (or~$\gd\wt\o,\tA,\tB,$)
generated by a \s ic metric tensor $\g_\(AB)$ (or~$\g_\(\tA\tB)$). Both \cn s are \LC \cn s. We get
\beq14.9
\bal
D\Psi&= d\Psi + \o^{AB}\wh\si_{AB}\Psi\\
D\ov\Psi&= d\ov\Psi - \ov\Psi\o^{AB}\wh\si_{AB}
\eal
\e
or
\beq14.10
\bal
D\Psi&= d\Psi + \o^{\tA\tB}\wh\si_{\tA\tB}\Psi\\
D\ov\Psi&= d\ov\Psi - \ov\Psi\o^{\tA\tB}\wh\si_{\tA\tB}
\eal
\e

Moreover, we consider in Refs \cite{ax4}, \cite{ax5} new kinds of ``gauge'' \dv s. We generalized this approach to an arbitrary gauge group~$G$
(or~$H$) (see Ref.~\cite{ax6})
\beq14.11
\bal
\cD \Psi &= \hor(D\Psi) = \gv d\Psi + \hor(\o^{AB})\wh\si_{AB}\Psi\\
\cD \ov\Psi &= \hor(D\ov\Psi) = \gv d\ov\Psi + \ov\Psi \hor(\o^{AB})\wh\si_{AB}
\eal
\e
or
\beq14.12
\bal
\cD \Psi &= \hor(D\Psi) = \gv d\Psi + \hor(\o^{\tA\tB})\wh\si_{\tA\tB}\Psi\\
\cD \ov\Psi &= \hor(D\ov\Psi) = \gv d\ov\Psi + \ov\Psi \hor(\o^{\tA\tB})\wh\si_{\tA\tB}
\eal
\e
Horizontality is understood in the sense of a \cn~$\o$ on the bundle $\ul P$.

It is easy to see that a \cn
\beq14.13
\wh\o_{AB} = \hor(\wt\o_{AB})
\e
defined by our new gauge \dv s is a metric \cn\ on~$P$, moreover, with a non-vanishing torsion. In this way we get a consistent theory of
\ci\ differentiation of spinor and tensor (vector) fields working with this \cn\ in place of $\wt\o_{AB}$. This remark is also applied for the~\cn
\beq14.14
\wh\o_{\tA\tB} = \hor(\wt\o_{\tA\tB})
\e
We refer also to Appendix A and Appendix B.

Let us notice the \fw\ fact. All the formalism considered here can be extended to Rarita--Schwinger field, i.e.\ to $3/2$-spin (not only to
$1/2$-spin) field. In order to do this we consider one-form spinor fields. It means, we consider on~$P$ horizontal 1-forms (horizontality is
understood in the sense of a \cn~$\o$ on a principal fiber bundle~$\ul P$). They take values in a \fn\ \rp ation of
$\SO(1,n+3)$ ($\Spin(1,n+3)$) or $\SO(1,n+n_1+3)$ ($\Spin(1,n+n_1+3)$) or $\SO(1,19)$ ($\Spin(1,19)$). We have
\bg 14.15
\Psi=\Psi_M\t^M, \qquad \ov\Psi=\ov\Psi_M\t^M, \qquad \ov\Psi_M=\Psi_M^+ \G^4\\
\Psi=\Psi_\mu\t^\mu, \qquad \ov\Psi=\ov\Psi_\mu\t^\mu, \qquad \ov\Psi_\mu=\Psi_\mu^+ \G^4. \lb{14.16}
\e
Eqs \er{14.15} correspond to the case with a \sn\ \s y breaking, and Eqs \er{14.16} to the case without a \sn\ \s y breaking (an absence
of a manifold $M=G/G_0$ or~$S^2$, see Ref.~\cite{aQ}).

We assume that these one-form spinor fields depend on the group \cd s in a trivial way, i.e.\ by the action of the group~$G$ ($H$~or~$G2$).
We introduce for these one-form spinor fields a new kind of gauge \dv s similarly as for a $0$-form spinor fields case
\bg 14.17
\cD\Psi = \hor(D\Psi) = d_1\Psi + \hor(\wt\o^{AB})\wh\si_{AB}\land \Psi \\
\cD\ov\Psi = \hor(D\ov\Psi) = d_1\ov\Psi - \hor(\wt\o^{AB})\land\ov\Psi \wh\si_{AB} \lb{14.18}
\e
where
\bg14.19
D\Psi = d\Psi + \wt\o{}^{AB} \wh\si_{AB}\land\Psi\\
D\ov\Psi = d\ov\Psi - \wt\o{}^{AB}\land\ov\Psi \wh\si_{AB} \lb{14.20} \\
d_1\Psi=\hor(d\Psi) \lb{14.21} \\
d_1\ov\Psi=\hor(d\ov\Psi) \label{14.22}
\e
are ordinary ``gauge'' \dv s on~$E$ (or on $E\tm M$) if we take in the place of $A,B$ also $\wt A,\wt B$. In this way we take similarly
\bg14.23
\cD\Psi = d_1\Psi + \hor(\wt\o{}^{\tA\tB})\wh\si_{\tA\tB}\land\Psi\\
\cD\ov\Psi = d_1\ov\Psi - \hor(\wt\o{}^{\tA\tB})\land\ov\Psi\wh\si_{\tA\tB}\label{14.24}
\e
It is understandable that we have in both cases Eq.~\er{14.4} for $1$-form spinor fields.

It is easy to see that we get
\bg14.25
\cD\Psi = \gv{\wt\cD}\Psi - \tfrac18 \la \gdg H,a,\g,b,[\G_a,\G_b] \t^\g \land \Psi \\
\cD\ov\Psi = \gv{\wt\cD}\ov\Psi + \tfrac18 \la \gdg H,a,\g,b, \t^\g \land \ov\Psi[\G_a,\G_b] , \lb{14.26}
\e
where
\bg14.27
\gv{\wt\cD}\Psi = \hor(\wt D \Psi)\\
\gv{\wt\cD}\ov\Psi = \hor(\wt D \ov\Psi). \lb{14.28}
\e
$\gv{\wt\cD}\Psi$ and $\gv{\wt\cD}\ov\Psi$ are exterior \ci\ \dv s on~$E$ (or $E\tm M=E\tm G/G_0$, or $E\tm S^2$)
\wrt a \LC \cn\ on~$E$ (or $E\tm M=E\tm G/G_0$, or $E\tm S^2$) and ``gauge'' at once.

\section{Lagrangians for \fe\ fields}
Let us define \Lg s for \fe\ fields defined on a manifold~$P$ in several cases (see Refs \cite{ay1}--\cite{ay10}). Let $\Psi(x,y)$ be
a spinor field in a \fn\ \rp ation of $\SO(1,n+3)$ ($\Spin(1,n+3)$) where $x\in E$, $y\in G$ (in a local trivialization). Let $\F_i(y)$ be
zero modes on~$G$, i.e.
\beq5.1
\Dint \F_i=0
\e
where
\beq5.2
\Dint=\G^a \pa_a, \q a=5,6,\dots,n+4.
\e
In this way we can write
\beq5.3
\Psi(x,y) = \sum_i \Psi_i(x)\F_i(y)
\e
and
\beq5.4
\Dint \Psi=0.
\e
It means we consider on $P$ only zero-modes \wrt $G$
\beq5.5
\ov\Psi(x,y) = \Psi^+(x,y)\G^4 = \sum_i \Psi_i^+ (x)\F_i^\ast (y)\G^4 = \sum_i\ov \Psi_i(x)\F_i^\ast(y).
\e
Our zero-modes on $G$ form a complete orthonormal basis
\beq5.6
\int_G d\mu(y)\, \F_i(y)\F_j^\ast(y) = \d_{ij}.
\e
In this way we define a \Lg\ for \fe s
\beq5.7
\cLf = \frac12\,i\hbar c\int_G d\mu_G(y)\Bigl(\ov\Psi\G^M \gv{\wt\n_M}\Psi - \gv{\wt\n_M} \ov\Psi\G^M\Psi\Bigr).
\e
$\mu_G$ is a bi\iv t measure on $G$.

Moreover, our \ci\ \dv\ (defined in Section \ref{s:spinor}) is \wrt a horizontal part of \LC \cn\ on~$P$ and one gets
(see Appendix A and B)
\bml5.8
\cLf = \frac12\,i\hbar c\sum_i\Bigl(\ov\Psi_i(x^\a)\G^\mu \gv{\wt{\ov\n}_\mu} \Psi_i(x^\a) - \gv{\wt{\ov \n}_\mu} \ov\Psi_i(x^\a)
\G^\mu\Psi_i(x^\a)\Bigr) \\
{}= \sum_i \cL_D (\Psi_i,\ov\Psi_i,\gv{\wt{\ov D}}) + i\,\frac{4\ell\pl}{\sqrt\a}\,qH^{a\a\g}h_{ab}
\ov\Psi_i \G^b[\G_\a,\G_\g]\Psi_i,
\e
$\cL_D(\Psi_i,\ov\Psi_i,\gv{\wt{\ov\n}})$ is an ordinary Dirac \Lg\ for spinor field~$\Psi_i$,
where $\gv{\wt{\ov D}}$ means an ordinary exterior \ci\ \dv\ \wrt \LC \cn\ on~$E$ and ``gauge'' at once. We use of course Eq.~\er{5.6}.
$\ell\pl$ means a Planck's length, $q$---an \el ary charge, $\a$~a~fine \sc e \ct. We get here some anomalous terms which can go to PC breaking
(see Ref.~\cite{q}). Let us come to more
complicated case where we have to do with a \sn\ \s y breaking and Higgs' field, i.e.\ we define \spf s on~$P$ which is now a bundle manifold
over $E\tm G/G_0$ with a \sc al group~$H$. In this case we have
\beq5.9
\Psi(x,x_1,y) = \sum_i \sum_k \Psi_{ik}(x^\a) \wt\Phi_i(x_1)\wh\Phi(y), \q x\in E,\ x_1\in M=G/G_0, \ y\in H,
\e
(in a local trivialization), where $\wh\Phi{}^k(y)$ are zero-modes on $H$,
\bg5.10
\int d\mu_H(y) \wh\Phi_k(y)\wh\Phi_l(y) = \d_{kl} \\
\Dint \wh\Phi_k =0. \lb{5.11}
\e
$\wh\Phi_i(x_1)$, $x_1\in M=G/G_0$ are not in general zero-modes on~$M$. Moreover, these \f s form an orthonormal set defined on~$M$,
\beq5.12
\int_M d\mu_M(x_1) \,\wt\Phi_1(x_1)\wh\Phi{}^\ast_j (x_1)=\d_{ij},
\e
$d\mu_H(y)$ and $d\mu_M(x_1)$ are measures on~$H$ and~$M$, respectively. $r$~is a radius of the manifold~$M$ which gives us a~scale of masses
of \fe s. Usually it is supposed that $\mu_H$ is a bi\iv t measure on~$H$.
\bg5.13
\ov\Psi = \G^4\Psi^+ = \sum_i \sum_k \ov\Psi_{ik}(x^\a)\wt\Phi{}^\ast(x_1)\wh\Phi{}^\ast_k(y)\\
\ov\Psi_{ik} = \G^4\Psi^+_{ik}. \lb{5.14}
\e
The \Lg\ for \fe s is defined  in this case as follows
\beq5.15
\cLf = \frac{i\hbar c}2 \int_M d\mu_M(x_1)\int d\mu_H(y)\Bigl(\ov\Psi \G^M \gv{\wt\n_M}\Psi- \gv{\wt\n_M}\ov\Psi \G^M\Psi\Bigr).
\e
One easily gets
\bml5.16
\cLf = \sum_i \sum_k \biggl(\cL_D(\Psi_{ik},\ov\Psi_{ik},\gv{\wt{\ov D}}) + i\,\frac{4\ell\pl q}{\sqrt\a}\,H^{a\a\g} h_{ab}\ov\Psi_{ik}\G^b
[\G_\a,\G_\g]\Psi_{ik}\\
{}+ \frac{i\hbar c}r\, \gd\Phi,a,\td m,\ov\Psi_{ik} \G^{\td m} \wh X_a \Psi_{ik} - \frac{\hbar c}{8r}\, \gv{\n_\b} \gd \ov\Phi,a,\td m,
\ov\Psi_{ik} \G^{\td m}[\G^b,\G^b] h_{ab}\Psi_{ik}\\
{}+ \frac{i\hbar c}{2r} \sum_j \ov\Psi_{ij}\G^{\td m} \wh\rho_{\td mkj}\Psi_{ik}\biggr)
\e
where
\beq5.17
\wh\rho_{\td mkj} = \int_M d\wt\mu_M(x_1) \bigl(\wt\Phi{}^\ast_j (x_1)\xi_{\td m}\wt\Phi_k(x_1) + \xi_{\td m}\wt\Phi^\ast_k(x_1)\wt\Phi_j(x_1)\bigl),
\e
$\wt\mu_M(x_1)$ is normalized measure on~$M$. It means we take $g_\(\td a\td b)$ (see Eq.~\er{4.16}).

$\wh X_a$ mean matrices of  $\fh$ algebra generators in a spinor \rp ation, $\xi_{\td a}$ vector fields acting along \cd\ lines on $M=G/G_0$,
$\cL_D(\Psi_{ik},\ov\Psi_{ik},\gv{\wt{\ov D}})$ is an ordinary Dirac \Lg\ for $\Psi_{ik}$.

$\gv{\wt{\ov D}}$ means an exterior \ci\ \dv\ \wrt a \LC \cn\ generated by $g_\(\a\b)$ on~$E$ and ``gauge'' at once.

It is easy to see that
\beq5.18
i\hbar c \sum_i \sum_k \gd\Phi,a,\td m, \ov\Psi_{ik} \G^{\td m}\wh X_a\Psi_{ik}
\e
is a Yukawa \ia\ term and can be a source of masses for \fe s after a \sn\ \s y breaking and \Hm.

The term
\beq5.19
\frac{i\hbar c}{2r} \sum_j \ov\Psi_{ij}\G^{\td m} \wh\rho_{\td mkj}\Psi_{ik}
\e
can be also a source of masses and mixing angles for \fe s. Moreover, the term
\beq5.20
i\,\frac{4\ell\pl q}{\sqrt\a}\,H^{a\a\g} h_{ab}\ov\Psi_{ik}\G^b [\G_\a,\G_\g]\Psi_{ik}
\e
defines an anomalous electric-like dipole \ia\ with \YM' field. We get here anomalous terms which can go to PC breaking (see Ref.~\cite{q}).

Let us come to the GSW model in our setting. In this case a \Lg\ for \fe s is defined:
\bml5.21
\cLf = \frac{i\hbar c}{2} \int_{S^2 \tm G2} \sin\psi\,d\psi\,d\vf\,d\mu_{G2}(y) \bigl(\ov\Psi\G^M \gv{\wt\n_M}\Psi - \gv{\wt\n_M}\ov \Psi
\G^M \Psi\bigr) \\
{}=\frac{i\hbar c}2 \sum_i \int_{S^2} \sin\psi\,d\psi\,d\vf \biggl(\sum_{\dwa{j,m\\|m|\le j}}\ov\Psi_{ijm}Y^\ast_{jm}(\psi,\vf)
\G^M \gv{\wt\n_M} \sum_{\dwa{l,n\\|n|\le l}}\ov\Psi_{iln}Y_{ln}(\psi,\vf)\\
{}- \gv{\wt\n_M} \sum_{\dwa{j,m\\|m|\le j}} \ov\Psi_{ijm}
Y^\ast_{jm}(\psi,\vf)\G^M \sum_{\dwa{l,n\\|n|\le l}}\ov\Psi_{iln}Y_{ln}(\psi,\vf)\biggr)
\e
where $Y_{ln}$ are spherical harmonics (see Refs \cite{az1}--\cite{az3}) and
\bg5.22
\Psi(x,\psi,\vf,y) = \sum_i \sum_{j,m} \Psi_{ijm}(x) Y_{jm}(\psi,\vf) \Phi_i(y)\\
\int_0^\pi \int_0^{2\pi} \sin\psi\,d\psi\,d\vf\,Y_{jm}(\psi,\vf)Y^\ast_{ln}(\psi,\vf) = \d_{jl}\d_{mn} \lb{5.23} \\
Y_{jm}(\psi,\vf) = e^{im\vf}\sqrt{\frac{(2j+1)(j-m)!}{4\pi(j+m)!}}\,P^m_l(\cos\psi), \lb{5.24}
\e
$j=0,1,2,\dots$, $m=-j,-j+1,\dots,j-1,j$,  $|m|\le j$. $\Phi_i(y)$ are zero-modes on $G2$ manifold.
\bg5.25
\ov\Psi(x^\a,\psi,\vf,y) = \sum_i \sum_{j,m} \ov\Psi_{ijm}(x^\a)Y^\ast_{jm}(\psi,\vf)\Phi^\ast_i(y)\\
\int_{G2} d\mu_{G2}(y)\,\Phi_i(y)\Phi^\ast_j(y) = \d_{ij}. \lb{5.26}
\e
They form an orthonormal basis on $G2$.

$P^m_l(x)$ are associated Legendre polynomials defined on $\langle-1,1\rangle$ intervals for $|m|\le l$. In this way we can proceed calculations:
\bml5.27
\cLf = \frac{i\hbar c}{2}\sum_i \Biggl\{\sum_{\dwa{j,m\\|m|\le j}} \sum_{\dwa{l,n\\|n|\le l}} \int_{S^2} \sin\psi \, d\psi\,d\vf \Bigl[
\bigl(\ov\Psi_{ijm}\G^\mu \gv{\wt \n_\mu}\Psi_{iln}Y^\ast_{jm}(\psi,\vf)Y_{ln}(\psi,\vf)\\
{}-\gv{\wt\n_\mu}\ov\Psi_{ilm}\G^\mu\Psi_{ijm}Y^\ast_{lm}(\psi,\vf)Y_{jn}(\psi,\vf)\bigr)
+\bigl(\ov\Psi_{ijm}\G^{\td m}\Psi_{iln}Y^\ast_{jm}(\psi,\vf)\xi_{\td m}Y_{ln}(\psi,\vf)\\
{}-\ov\Psi_{ijm}\G^{\td m}\Psi_{iln}\xi_{\td m} Y^\ast_{jm}(\psi,\vf)Y_{ln}(\psi,\vf)\bigr)\Bigr]\\
{}=\frac{i\hbar c}2 \sum_i \sum_{\dwa{j,m\\|m|\le j}}\bigl(\ov\Psi_{ijm}\G^\mu\gv{\wt\n_\mu}\Psi_{ijm}
-\gv{\wt\n_\mu}\ov\Psi_{ijm}\G^\mu\Psi_{ijm}\bigr)\\
{}+\int_0^\pi \int_0^{2\pi} \sin\psi\,d\psi\,d\vf \sum_{\dwa{j,m\\|m|\le j}} \sum_{\dwa{l,n\\|n|\le l}}
\biggl[\biggl(\ov\Psi_{ijm}\G^5 \bigl(\vf^\ast_1 x_{-\a}+\vf^\ast_2x_{-\b} - \vf_1x_\a-\vf_2x_\b\bigr)
\Psi_{iln}Y^\ast_{jm}(\psi,\vf)\\
{}- i\sin\psi\ \ov\Psi_{ijm}\G^6 \bigl(\vf_1x_\a + \vf_2x_\b + \vf_1^\ast x_{-\a}+\vf_2^\ast x_\b\bigr)\Psi_{iln}
Y^\ast_{jm}(\psi,\vf) Y_{ln}(\psi,\vf)\\
{}-i\cos\psi \ov\Psi_{ijm}\G^6 (h_\a+h_\b)\Psi_{iln}Y^\ast_{jm}(\psi,\vf)Y_{ln}(\psi,\vf)\\
{}+ \frac i4 \Bigl[\ov\Psi_{ijm}\Bigl(\pa_\mu\vf_1 - \frac12\,iA_\mu^- \vf_2
-\frac 12\,iA^3_\mu \vf_1 - \frac{\sqrt3\,i}6\, B_\mu\vf_1\Bigr)x_\a h^{x_\a a}\\
{} + \Bigl(\pa_\mu\vf_1 -\frac12\,iA^+_\mu\vf_1 - \frac{\sqrt3}6\, iB_\mu\vf^\ast_2\Bigr)x_{-\a}h^{x_{-\a}a}
+\Bigl(\pa_\mu\vf_1 - \frac12\,iA^+_\mu \vf_1+ \frac12\,iA^+_\mu\vf_2 - \frac{\sqrt3}6\,i B_\mu\vf_2\Bigr)x_\b h^{x_\b a}\\
{}-\Bigl(\pa_\mu\vf_2^\ast + \frac12\,iA^- _\mu \vf_1^\ast -\frac12\,A_\mu^3 \vf^\ast_2
+\frac{\sqrt3}6\, i B_\mu\vf_2^\ast\Bigr) x_{-\b}h^{x_{-\b a}}\Bigr] \G^5 [\G^\mu,\G_a] \Psi_{iln}Y_{jm}^\ast(\psi,\vf)Y_{ln}(\psi,\vf)\\
{}- i\sin\psi \ov\Psi_{ijm} \Bigl[\Bigl(\pa_\mu\vf_1 - \frac12\,iA^+_\mu \vf_2 - \frac12\,A^3_\mu\vf_1 - \frac{\sqrt3}6\,i B_\mu\vf_1\Bigr)
x_\a h^{x_\a a}\\
{}+\Bigl(\pa_\mu\vf_2 - \frac12\,iA^+_\mu\vf_2 - \frac{\sqrt3}6\,i B_\mu\vf_2\Bigr)x_\b h^{x_\b a}
-\Bigl(\pa_\mu\vf^\ast_1 + \frac12\,i A^3_\mu\vf_1^\ast + \frac{\sqrt3}6\,iB_\mu \vf_1^\ast\Bigr)x_{-\a}h^{x_{-a}a}\\
{}-\Bigl(\pa_\mu\vf_2^\ast + \frac12\,A^-_\mu\vf_1^\ast - \frac12\,iA^3_\mu\vf_2^\ast + i\,\frac{\sqrt3}6\,B_\mu\vf^\ast_2\Bigr)x_{-\b}h^{x_{-\b}a}\Bigr]
\G^6 [\G^\mu,\G^a]\Psi_{iln}\biggr) Y^\ast_{jm}(\psi,\vf)Y_{ln}(\psi,\vf)\biggr]\\
{}+\Bigl\{\ov\Psi_{ijm}\G^5_\(in)\Psi_{iln}Y_{jm}^\ast(\psi,\vf)Y_{ln}(\psi,\vf)+ \frac{i}8\sin2\psi\,\ov\Psi_{ijm}(\G^5+3\G^6)
\Psi_{iln}Y_{jm}^\ast(\psi,\vf) Y_{ln}(\psi,\vf)\\
{} + \ov\Psi_{ijm}\G^5\Psi_{iln}Y_{jm}^\ast(\psi,\vf)\frac{\pa}{\pa\psi}Y_{ln}(\psi,\vf)\Bigr\}\Biggr\},
\e
a continuation of Eq.~\er{5.21}.

Let us notice the \fw\ facts:
\begin{gather}
-i\int_0^\pi \int_0^{2\pi} \sin^2\psi \, Y^\ast_{jm}(\psi,\vf) Y_{ln}(\psi,\vf)\,d\psi\,d\vf \hskip200pt \nn \\
\hskip60pt {}= -i\,\frac{\d_{mn}}{4\pi} \sqrt{\frac{(2l+1)(2j+1)(l-m)!(j-m)!}{(l+m)!(j+m)!}}
\int_{-1}^1 dx\,(1-x^2)^{1/2} P^m_l(x)P^m_j(x) \lb{5.28} \\
-i \int_0^\pi \int_0^{2\pi} \sin\psi \cos\psi\, Y^\ast_{jm}(\psi,\vf)Y_{ln}(\psi,\vf)\,d\psi\,d\vf \hskip200pt \nn\\
\hskip60pt {}= -i\,\frac{\d_{mn}}{4\pi} \sqrt{\frac{(2l+1)(2j+1)(l-m)!(j-m)!}{(l+m)!(j+m)!}}
\int_{-1}^1 dx\,x P^m_l(x)P^m_j(x) \lb{5.29} \\
\int_0^\pi \int_0^{2\pi} \sin\psi\, Y^\ast_{jm}(\psi,\vf) \,\frac{\pa}{\pa \psi}Y_{ln}(\psi,\vf) \,d\psi\,d\vf \hskip200pt \nn \\
{} = -\frac{\d_{mn}}{4\pi} \sqrt{\frac{(2l+1)(2j+1)(l-m)!(j-m)!}{(l+m)!(j+m)!}}\hskip100pt \nn \\
\hskip50pt {}\tm\biggl[(m+1)\int_{-1}^1 \frac{xP^m_j(x)P^m_l(x)}{\sqrt{1-x^2}}\,dx + (l-m-1)\int_{-1}^1 \frac{xP^m_j(x)P^{m+1}_l(x)}{\sqrt{1-x^2}}\,dx\biggr] \lb{5.30} \\
\int_0^\pi \int_0^{2\pi} \sin\psi\, Y^\ast_{jm}(\psi,\vf)\,\pap{}\psi Y_{ln}(\psi,\vf)\,d\psi\,d\vf =
in\d_{jl}\d_{mn} \lb{5.31} \\
\int_0^\pi \int_0^{2\pi} \sin^2\psi \cos\psi\, Y^\ast_{jm}(\psi,\vf)Y_{ln}(\psi,\vf)\,d\psi\,d\vf \hskip200pt \nn \\
\hskip60pt {}= \frac{\d_{mn}}4 \sqrt{\frac{(2l+1)(2j+1)(l-m)!(j-m)!}{(l+m)!(j+m)!}}
\int_{-1}^1 dx\,x(1-x^2)^{1/2} P^m_j(x)P^m_l(x) \lb{5.32}
\end{gather}

By using Eqs \er{5.28}--\er{5.32} our \Lg\ for \fe s starts to be more simple
\bml5.33
\cLf = \frac{i\hbar c}2 \sum_i \Biggl\{\sum_{j,m} \Bigl[ \bigl(\ov\Psi_{ijm}\G^\mu \gv{\wt{\ov\n}_\mu}\Psi_{ijm}
- \gv{\wt{\ov\n}_\mu} \ov\Psi_{ijm}\G^\mu\Psi_{ijm}\bigr)\\
{}+ \ov\Psi_{ijm}\G^5\bigl(\vf^\ast_1 x_{-\a}+\vf^\ast_2x_{-\b} - \vf_1x_\a-\vf_2x_\b\bigr)\Psi_{ijm}\Bigr]\\
{}+ \sum_{j,l,m} \Bigl\{ -\frac i{4\pi}\ov\Psi_{ijm} \G^6\bigl(\vf_1x_\a+\vf_2x_\b + \vf^\ast_1x_{-\a}+\vf^\ast_2x_\b\bigr)\Psi_{ijm}
\sqrt{\frac{(2l+1)(2j+1)(l-m)!(j-m)!}{(l+m)!(j+m)!}}\\
\hskip60pt{}\tm\int_{-1}^1 dx\, (1-x^2)^{1/2}P^m_l(x)P_j^m(x)\\
{}-\frac i{4\pi} \ov\Psi_{ijm}\G^6(h_\a+h_\b)\Psi_{ijm}
\sqrt{\frac{(2l+1)(2j+1)(l-m)!(j-m)!}{(l+m)!(j+m)!}}
\int_{-1}^1 dx\,xP^m_l(x)P^m_j(x)\Bigr\}\\
{}\frac i4 \biggl[\sum_{j,m}\ov\Psi_{ijm}\Bigl(\pa_\mu\vf_1 - \frac 12\,iA^-_\mu\vf_2 - \frac12\,iA^3_\mu\vf_1 - \frac{\sqrt3\,i}{6}\,B_\mu\vf_1\Bigr)
x_\a h^{x_\a a}\\
{}+\Bigl(\pa_\mu\vf_1 - \frac i2\,A^+_\mu\vf_1 - \frac{\sqrt3}6 \,iB_\mu\vf_2^\ast\Bigr)x_{-\a}h^{x_{-\a}a}
-\Bigl(\pa_\mu\vf^\ast\vf_2 + \frac i2\, A^-\vf_1^\ast - \frac12\, A^3_\mu\vf_2^\ast + \frac{\sqrt3}6 \,iB_\mu\vf_2^\ast\Bigr)x_{-\b} h^{x_{-\b}a}\\
{}+\Bigl(\pa_\mu\vf_1 - \frac12\,iA^+_\mu\vf_1 + \frac12\,i A_\mu^+\vf_2 - \frac{\sqrt3}6\,iB_\mu\vf_2\Bigr)x_\b h^{x_\b a}\biggr]\G^5[\G^\mu,\G_a]\\
{}-\frac i{4\pi} \sum_{j,l,m} \biggl\{\Psi_{ijm}
\biggl[\Bigl(\pa_\mu\vf_1 - \frac12\,A^-_\mu\vf_2 - \frac i2\,A^3_\mu\vf_1-\frac{\sqrt3}6\,iB_\mu\vf_1\Bigr)x_\a h^{x_\a a}\\
{}+\Bigl(\pa_\mu\vf_2 - \frac i2\,A^+_\mu\vf_2 - \frac{\sqrt3}6\,B_\mu\vf_2\Bigr)x_\b h^{x_\b a}
-\Bigl(\pa_\mu\vf^\ast_1+ \frac12\,iA^3_\mu\vf_1^\ast + \frac{\sqrt3}6\,i\Bm\vf_1^\ast\Bigr)x_{-\a}h^{x_{-\a}a}\\
{}-\Bigl(\pa_\mu\vf_2^\ast + \frac12\,\Am^-\vf_1^\ast -\frac12\,\Am^3\vf_2^\ast+i\,\frac{\sqrt3}6\,\Bm\vf_2^\ast\Bigr)x_{-\b}h^{x_{-b}a}\biggr]
\Psi_{ilm}\sqrt{\frac{(2l+1)(2j+1)(l-m)!(j-m)!}{(l+m)!(j+m)!}}\\
\hskip80pt{}\tm\int_{-1}^1 dx\,x P^m_l(x)P^m_j(x) \\
{}+ \frac{1}4 \ov\Psi_{ijm}(\G^5+3\G^6)\Psi_{ilm}\sqrt{\frac{(2l+1)(2j+1)(l-m)!(j-m)!}{(l+m)!(j+m)!}}
\int_{-1}^1 dx\,x P^m_l(x)P^m_j(x)\biggr\}\\
{}+i \sum_{j,m} m \ov\Psi_{ijm}\G^5 \Psi_{ijm}
-\frac1{4\pi}\sum_{j,l,m}\ov\Psi_{ijm}\G^5 \Psi_{ilm}\sqrt{\frac{(2l+1)(2j+1)(l-m)!(j-m)!}{(l+m)!(j+m)!}}\\
{}\tm\biggl[(m+1)\int_{-1}^1 \frac{xP^m_j(x)P^m_l(x)}{\sqrt{1-x^2}}\,dx + (l-m-1)\int_{-1}^1 \frac{P^m_j(x)P^{m+1}l(x)}{\sqrt{1-x^2}}\,dx
\biggr]\Biggr\}
\e

In this \Lg\ we have a Yukawa \ia\ which after a \ssb\ and \Hm\ results in masses generation of \fe s and possible mixing. These are the terms:

\medskip
\leftline{$1^\circ$ $\displaystyle \frac{i\hbar c}{2r} \sum_i \sum_{j,m} \Psi_{ijm}\G^5\bigl(\vf_1^\ast x_{-\a}+\vf_2^\ast x_{-\b} - \vf_1x_\a - \vf_2x_\b\bigr)
\Psi_{ijm}$}

\leftline{$2^\circ$ $\displaystyle \frac{i\hbar c}{2r} \sum_{i,j,l,m}\biggl\{-\frac1{4\pi} \ov\Psi_{ijm}\G^6 \bigl(\vf_1 x_{\a}+\vf_2 x_{\b}
+ \vf_2^\ast x_{-\a} + \vf_2^\ast x_\b\bigr) \Psi_{ijm}$}
\centerline{$\displaystyle \sqrt{\frac{(2l+1)(2j+1)(l-m)!(j-m)!}{(l+m)!(j+m)!}} \int_{-1}^1 dx\,
(1-x^2)^{1/2} P^m_l(x)P^m_j(x)$}
\rightline{$\displaystyle -\frac1{4\pi} \ov\Psi_{ijm}\G^6(h_\a+h_\b)\Psi_{ilm}\sqrt{\frac{(2l+1)(2j+1)(l-m)!(j-m)!}{(l+m)!(j+m)!}}
\int_{-1}^1 dx\,xP^m_l(x)P^m_j(x)\biggr\}$}
\medskip

We have some additional terms which can generate masses and mixing angles

\medskip
\leftline{$1^\circ$ $\displaystyle -\frac{\hbar c}{2r} \sum_i \sum_m \Psi_{ijm}\G^5 \Psi_{ijm}$}

\leftline{$2^\circ$ $\displaystyle -\frac{i\hbar c}{4\pi r} \sum_i \sum_{j,l} \ov\Psi_{ijm}\G^5 \Psi_{ilm}\sqrt{\frac{(2l+1)(2j+1)(l-m)!(j-m)!}{(l+m)!(j+m)!}}$}
\rightline{$\displaystyle {}\tm\biggl[(m+1)\int_{-1}^1 \frac{x P^m_j(x)P^m_l(x)}{\sqrt{1-x^2}}\,dx + (l-m-1)\int_{-1}^1 \frac{P^m_j(x)P^{m+1}_l}{\sqrt{1-x^2}}\,dx
\biggr]$}

\leftline{$3^\circ$ $\displaystyle -\frac{\hbar c}{16\pi r^2} \sum_{i,j,m}\ov\Psi_{ijm}(\G^5+3\G^6)\Psi_{ilm}
\sqrt{\frac{(2l+1)(2j+1)(l-m)!(j-m)!}{(l+m)!(j+m)!}}\int_{-1}^1 dx\, P_l^m(x)P_j^m(x)$,}

\noindent $r$ is a radius of a sphere $S^2$ which gives a scale of masses for \fe s.

Thus we have many possibilities in a \fe\ sector, even if we remove from the theory \fe\ with \Pl's masses via zero-mode condition. In all the formulas written above
$$
|m|\le \min(j,l).
$$

The important problem in the theory of \fe s is to get chiral \fe s. We are supposing of course that the 6-\di al (in general $(n_1+4)$-\di al) mass is zero,
i.e.
\beq5.34
\Dint\Psi=0
\e
(i.e. Eq. \er{5.2}). Otherwise the mass of a \fe\ is of \Pl's mass
\beq5.35
\Dint \Psi=\ov m\Psi.
\e
$\ov m\ne0$ is of order of \Pl's mass (see Refs \cite{ay1}--\cite{ay10}). Thus we have
$$
\Dint = \sum_{A=7}^{20}\G^A \partial_A.
$$
We demand zero-modes condition. One can write
\bg5.36
\wh\G{}^5=\g^5\ot\G^1, \q \wh\G{}^6 = \g^5\ot \G^2, \\
{\hbox to6pt{\rm D\hss}\hskip-5pt\slash}{}^4 = (\g^\mu\ot I)\pa_\mu, \lb{5.37} \\
\overset{(5,6)}{\hbox to6pt{\rm D\hss}\hskip-5pt\slash}=\g^5 \ot \Bigl(\G^1\,\pap{}\psi + \G^2\,\pap{}\vf\Bigr) \lb{5.38}
\e
and finally
\beq5.39
\Dint\Psi=\g^5 \ot \sum_{r=3}^{16} \G^r \pa_{r+4}\Psi.
\e
If
\bg5.40
\Dint \Phi_i(y)=0, \\
\Psi=\sum_i\Psi_i(x^\mu,\psi,\vf)\F_i(y) \lb{5.41}
\e
then
\beq5.42
N_c(\Dint)=(n_c^+ - n_c^- - n_{\bar c}^+ + n_{\bar c}^- )f.
\e
$n_c^+$ is a number of zero-modes of $\Dint$ for the Weyl spinors $\Psi^+$ associate with a complex \rp ation of the spinor, while $n_c^-$ and~$n_{\bar c}^+$
denote the corresponding values for $\Psi^-$ and for the complex conjugate \rp ation. $\Psi^+$~and~$\Psi^-$ Weyl spinors are defined as usual
$$
\bal
\Psi^\pm &= \frac12(1\pm \ov\G{}^{n+5})\Psi,\\
\Psi^\pm &= \frac12(1\pm \ov\G{}^{n+n_1+5})\Psi,\\
\Psi^\pm &= \frac12(1\pm \ov\G{}^{21})\Psi,
\eal
$$
where $\ov\G{}^{n+5}$, $\ov\G{}^{n+n_1+5}$, $\ov\G{}^{21}$ are higher \di al analogues of $\g^5$ matrix in four \di s (see Appendix~A). It is understandable
that we get left-handed spinors on~$E$ from the expansion~$\Psi^+$ and right-handed from the expansion of~$\Psi^-$. It means that we get $\Psi^+_{i}$ and
$\Psi^-_{i}$, $\Psi^+_{ik}$, $\Psi^-_{ik}$ or $\Psi^+_{ilm}$, $\Psi^-_{ilm}$.
The most important case because of physical applications is of course GSW model. It means $\Psi^+_{ilm}$, $\Psi^-_{ilm}$ and $\G^{21}$ matrix.
Thus $N_C(\Dint)$ is a number of 6-\di al left-handed \fe\ generations up to numerical factor, i.e.\ $f$ (see Appendix~C). Let $d_C$ be a \di\ of
a complex \rp ation~$C$.

We define
\beq5.43
N=\sum_C d_C|N_C(\Dint)|.
\e
We see that the total number of massless \fe s on $G2$ is
\beq5.44
n_0(G2)\ge N.
\e

Let us consider the \fw\ problem. Our \fe s are labeled by several indices, i.e.\ $i$, $ik$, $ilm$ and so on. Let us consider them as labeled by one index
$I$, i.e.\ $\Psi_I$, in particular $I=i$, $I=(i,k)$, $I=(i,l,m)$. Our spinors form of course a tower in many cases, this is an infinite tower. Moreover,
$\Psi_I$~belong to the \fn\ \rp ation of a group $\SO(1,n+3)$ ($\Spin(1,n+3)$), $\SO(1,n+n_1+3)$ ($\Spin(1,n+n_1+3)$) or in the case of GSW to $\SO(1,19)$
($\Spin(1,19)$).

We should decompose this \rp ation to a \rp ation of Lorentz group $\SO(1,3)$\break (${\rm SL}(2,\C)=\Spin(1,3)$). One gets
\bg5.45
D^F|_{\SO(1,3)}(\La) = L(\La)\oplus \ldots \oplus L(\La), \q \La\in\SO(1,3)\\
L(\La)=D^{(1/2,0)}(\La) \oplus D^{(0,1/2)} \lb{5.46}
\e
(see Ref.~\cite{ax3}).
\bg5.47
\Psi_I|_{\SO(1,3)} = \left[\begin{matrix}
\Psi_{I1} \\ \vdots \\ \Psi_{I2^{[n/2]}}
\end{matrix} \right]\\
\intertext{or}
\Psi_I|_{\SO(1,3)} = \left[\begin{matrix}
\Psi_{I1} \\ \vdots \\ \Psi_{I2^{[(n+n_1)/2]}}
\end{matrix} \right] \lb{5.48} \\
\Psi_I|_{\SO(1,3)} = \left[\begin{matrix}
\Psi_{I1} \\ \vdots \\ \Psi_{I2^8}
\end{matrix} \right] \lb{5.49}
\e
After taking a section of a bundle we get in any case
\beq5.50
e^*(\Psi_I)|_{\SO(1,3)} = \left[\begin{matrix}
\Psi_{I1} \\ \vdots \\ \Psi_{I2^N}
\end{matrix} \right]
\e
where $N=\[\frac n2]$ or $[\frac{n+n_1}2]$ or 8.

In the case of Weyl spinors we have similarly
\beq5.51
\bga
\Psi^+_{Ji}, \q i=1,2,\dots,2^{N-1}\\
\Psi^-_{Ji}, \q i=1,2,\dots,2^{N-1},
\ega
\e
where as usual
\beq5.52
e^*(\Psi^\pm_I)|_{\SO(1,3)} = \left[\begin{matrix}
\Psi^\pm_{I1} \\ \vdots \\ \Psi^\pm_{I2^{N-1}}
\end{matrix} \right].
\e

According to Ref.~\cite{aQ} we consider $\frac32$-spin spinor fields as $1$-forms defined on~$P$ horizontal \wrt a \cn\ defined on a fiber
bundle~$\ul P$,
$$
\hor\Psi=\Psi.
$$
In this way $\frac32$-spinor (-fields) are represented by spinor-valued forms (-fields). We have three cases as for ordinary $\frac12$-spinor
fields, i.e.\ $\ul P$ over~$E$ (\spt), $\ul P$~over $E\tm M=E\tm G/G_0$ and $\ul P$ over $E\tm S^2$. Moreover, we develop in details only the first case.

It is easy to see that we can proceed the same expansion procedure for $1$-form \spf s on a manifold~$P$ as for a $0$-form \spf s (i.e.\ \spf s).
One gets
\beq5.53
\Psi(x,y) = \sum_i \Psi_i(x) \Phi_i(y)
\e
where $\Psi_i(x)$ are now $1$-form \spf s
\beq5.54
\ov\Psi(x,y) = \sum_i \ov\Psi_i(x) \ov\Phi{}^\ast_i(y), \q x\in E,\ y\in G.
\e
Similarly one finds
\bea5.55
\Psi(x,x_1,y) &= \sum_i \sum_k \Psi_{ik}(x) \wt\Phi_i(x_1) \wh\Phi_k(y), \q x\in E,\ x_1\in M=G/G_0,\ y\in H,\\
\ov\Psi(x,x_1,y) &= \sum_i \sum_k \ov\Psi_{ik}(x) \wt\Phi{}^\ast_i(x_1) \wh\Phi{}^\ast_k(y), \lb{5.56}
\e
where as usual
\bea5.57
\ov\Psi _{ik}(x)&= \G^4 \Psi^+_{ik}(x) \\
\ov\Psi _{i}(x)&= \G^4 \Psi^+_{i}(x) \lb{5.58}
\e
and also
\bea5.59
\Psi(x,\psi,\vf,y) = \sum_i \sum_{\dwa{j,m\\|m|<j}} \Psi_{ijm}(x) Y_{jm}(\psi,\vf)\F_i(y) \\
\ov\Psi(x,\psi,\vf,y) = \sum_i \sum_{\dwa{j,m\\|m|<j}} \ov\Psi_{ijm}(x) Y^\ast_{jm}(\psi,\vf)\F^\ast_i(y) \lb{5.60}
\e
where $x\in E$, $\psi,\vf$ are $S^2$ \cd s, $y\in G2$ and
\beq5.61
\ov\Psi_{ijm} = \G^4 \Psi^+ _{ijm}.
\e

One can get \di al reduction procedure similarly as for \spf s finding
\beq5.62
e^*(\Psi_I)|_{\SO(1,3)} = \left[\begin{matrix}
\Psi_{I1} \\ \vdots \\ \Psi_{I2^N}
\end{matrix} \right]
\e
where $N=\[\frac n2]$ or $[\frac{n+n_1}2]$ or 8. But now $\Psi_{iI}$ are $1$-form \spf s on $E$ or $E\tm M=E\tm G/G_0$, or $E\tm S^2$.

One defines the \Lg s for Rarita--Schwinger fields as follows:
\beq5.63
\cL_{\rm Rarita-Schwinger} = \frac{i\hbar c}2 \int_M d\mu_M(x_1) \int_H d\mu_H(y)
\bigl(\ov\Psi \land \G_{2l+5}\G \land \cD\Psi - \cD\ov\Psi \land \G_{2l+5}\G \land \Psi\bigr)
\e
where
\beq5.64
\bal
\G&=\G_M \t^M\\
\Psi&=\Psi_M \t^M\\
\ov\Psi&=\ov\Psi_M \t^M
\eal
\e
and we impose supplementary conditions
\bg5.65
l\land \Psi = \ov\Psi \land l=0 \\
l= \G_M \eta^M \lb{5.66}
\e
where $\eta^ M$ is a dual Cartan base for $\t^M$. In the case of 20-\di al model (GSW~model) we have $S^2$ in the place of~$M$, and $H=G2$. In the case without
a \ssb\ we have only an integration for~$G$ (which replaces~$H$). We will not develop this formalism in the paper. This will be done elsewhere.

In the case without \s y breaking (an absence of $M$ or~$S^2$) one simply gets in the place of Eqs \er{5.64}--\er{5.66}
\bg5.67
\bal
\G&=\G_\mu \t^\mu\\
\Psi&=\Psi_\mu \t^\mu\\
\ov\Psi&=\ov\Psi_\mu \t^\mu
\eal \\
l\land \Psi = \ov\Psi \land l =0 \lb{5.68} \\
l=\G_\mu \eta^\mu \lb{5.69}
\e
where $\eta^\mu$ is a dual Cartan base for $\t^\mu$.

It is easy to see that in the case with a \ssb\ we have in the theory also a multiplet of spinor $1/2$-fields $\psi_a$, $a=5,6,\dots,n_1+4$ (or~$a=5,6$).

In the first case the Rarita--Schwinger field \Lg\ looks
\bml5.71
\cL_{\text{Rarita--Schwinger}}=\frac12\, i\hbar c\int_G d\mu_G \bigl(\ov\Psi \land \G_{2l+5}\G\land\cD \Psi - \cD\ov\Psi \land \G_{2l+5}
\G\land\Psi\bigr)\\
{}=\sum_j \cL_{\text{Rarita--Schwinger}}(\Psi_j,\ov\Psi_j,\wt\cD)+ \frac{i\hbar c\la}{2}
\sum_j \biggl[H^{\a\nu b}\Psi_{j\la}\G_b[\G_a,\G_\nu]\dg\Psi,j,\la,\\ + 2H^{\a\rho b}
\bigl(\ov\Psi_{j\la}\G_b\G^\la \G_\a \Psi_{j\rho} + \ov\Psi_{j\rho}\G_b\G^\la \G_\a\Psi_{j\la}
-\ov\Psi_{j\la}\G^\la\G_\rho\G_b\G^\nu\G_\a\Psi_{j\nu}\bigr)\biggr]\eta
\e
where
\beq5.72a
\cL_{\text{Rarita--Schwinger}}(\Psi_j,\ov\Psi_j,\wt\cD) = \frac{i\hbar c}2 \bigl(\ov\Psi_j\G_{2l+5}\land\G \land \wt\cD \Psi_j -
\wt\cD \ov\Psi_j\land\G_{2l+5}\land\G\land\Psi_j\bigr).
\e

Moreover, to give a taste of the full formalism we calculate the first terms in the \Lg\ involving \ia\ of $\frac32$-spin field and also
$\frac12$-spin field with \YM' and Higgs'
fields. Thus starting from Eq.~\eqref{5.63} we get as the first \ia\ term \er{5.72} in the \Lg\ development. The zero term is as usual (a~sum
of all fields free \Lg s)
$$
\sum_{ij}\cL_{\text{Rarita--Schwinger}}(\Psi_{ij},\ov\Psi_{ij},\wt\cD) + \sum_{i,j,\td l}\cL_{\text{Dirac}}(\Psi_{ij\td l},\ov\Psi_{ij\td l},\wt\cD).
$$
\begin{gather}
i\,\frac{\sqrt{G_N}\,\hbar}{4c} \sum_{i,j}\biggl\{ \a^b_k\biggl[\wt H{}^{\a\nu k}\Bigl(\ov\Psi_{ij\la}\G_b[\G_\a,\G_\nu]\dg\Psi,ij,\la,
+\ov\Psi_{ij\td l}\G_b[\G_\a,\G_\nu]\dg\Psi,ij,\td l,\Bigr)\nn\\
{}+2\wt H{}^{\a\rho k}\Bigl[\bigl(\ov\Psi_{ij\la}\G_b\G^\la \G_\a\Psi_{ij\rho}+\ov\Psi_{ij\td l}\G_b\G^{\td l}\G_\a\Psi_{ij\rho}\bigr)
+\bigl(\ov\Psi_{ij\rho}\G_b\G^\la\G_\a\Psi_{ij\la}+\ov\Psi_{ij\rho}\G_b\G^{\td l}\G_\a\Psi_{ij\td l}\bigr)\nn\\
{}-\Bigl(\ov\Psi_{ij\la}\G^\la\G_\rho\G_b\G^\nu\G_\a\Psi_{ij\nu} + \ov\Psi_{ij\td l}\G^{\td l}\G_\rho\G_b\G^\nu \G_\a\Psi_{ij\nu}\nn\\
{}+\ov\Psi_{ij\la}\G^\la\G_\rho\G_b\G^{\td n}\G_\a\Psi_{ij\td n} + \ov\Psi_{ij\td l}\G^{\td l}\G_\rho\G_b\G^{\td n}\G_\a\Psi_{ij\td n}\Bigr)\Bigr]\biggr]\nn\\
{}+\gv\n{}^\a\Phi^{b\td n}\bigl(\ov\Psi_{ij\la}\G_b[\G_\a,\G_{\td n}]\dg\Psi,ij,\la, + \ov\Psi_{ij\td l}\G_b[\G_\a,\G_{\td n}]\dg\Psi,ij,\td l,\bigr)\nn\\
{}+2\gv\n{}^\a \Phi^{b\td r}\Bigl[\bigl(\ov\Psi_{ij\la}\G_b\G^\la\G_\a\Psi_{ij\td r} + \ov\Psi_{ij\td l}\G_b\G^{\td l}\G_\a\Psi_{ij\td r}\bigr)
+\bigl(\ov\Psi_{ij\td r}\G_b\G^\la\G_\a\Psi_{ij\la}+\ov\Psi_{ij\td r}\G_b\G^{\td l}\G_\a\Psi_{ij\td l}\bigr)\nn\\
{}-\bigl(\ov\Psi_{ij\la}\G^\la\G_{\td r}\G_b\G^\nu\G_\a\Psi_{ij\nu}+ \ov\Psi_{ij\td l}\G^{\td l}\G_{\td r}\G_b\G^\nu\G_\a\Psi_{ij\nu}\nn\\
{}+\ov\Psi_{ij\la}\G^\la\G_{\td r}\G_b\G^{\td n}\G_\a \Psi_{ij\td n} + \ov\Psi_{ij\td l}\G^{\td l}\G_{\td r}\G_b\G^{\td n}\G_\a\Psi_{ij\td n}\bigr)\Bigr]\nn\\
{}+H^{b\td a\td n}\bigl(\ov\Psi_{ij\la}\G_b[\G_{\td a},\G_{\td n}]\dg\Psi,ij,\la, + \ov\Psi_{ij\td l}\G_b[\G_{\td a},\G_{\td n}]\dg\Psi,ij,\td l,)\nn\\
{}+2H^{b\td a\td r}\Bigl[\ov\Psi_{ij\la}\bigl(\G_b\G^\la\G_{\td a}\Psi_{ij\td r}+ \ov\Psi_{ij\td l}\G_b\G^{\td l}\G_{\td a}\Psi_{ij\td r}\bigr)
+\bigl(\ov\Psi_{ij\td r}\G_b\G^\la\G_\ta\Psi_{ij\la}+\ov\Psi_{ij\td r}\G_b\G^{\td l}\G_\ta\Psi_{ij\td l}\bigr)\nn\\
{}-\bigl(\ov\Psi_{ij\la}\G^\la\G_{\td r}\G_b\G^\nu\G^\ta\Psi_{ij\nu} + \ov\Psi_{ij\la}\G^\la\G_{\td r}\G_b\G^{\td n}\G_\ta\Psi_{ij\td n}\nn\\
{}+\ov\Psi_{ij\td l}\G^{\td l}\G_{\td r}\G_b\G^\nu\G_\ta\Psi_{ij\nu} + \ov\Psi_{ij\td l}\G^{\td l}\G_{\td r}\G_b\G^{\td n}\G_\ta\Psi_{ij\td n}\bigr)\Bigr]
\biggr\}\eta \lb5.72
\end{gather}
where
\beq5.73
H^{b\ta\td n}=\wt g{}^{(\ta\tc)}\wt g{}^{(\td n\td m)}\gd H,b,\tc\td m,.
\e
$\wt g^{(\ta\td n)}$ is an inverse tensor of $g_\(\ta\tb)$ on $M=G/G_0$, $\eta$~is a volume \el\ on $E\tm M$, $\Psi_{ij\nu}$ are $\frac32$-spin fields and
$\Psi_{ij\td n}$ are $\frac12$-spin fields.

In order to develop a formalism we should introduce several types of forms
\bea5.74
&\int_M d\mu_M(x_1)\t^{\td l}\wh\Phi_i(x_1)\wh\Phi{}^\ast_j(x_1)=\gd K,\td l,ij, & \hbox{1-forms} \\
&\int_M d\mu_M(x_1)\t^{\td l}\land \t^{\td n}\wh\Phi_i(x_1)\wh\Phi{}^\ast_j(x_1)=\gd L,\td l\td n,ij, & \hbox{2-forms} \lb 5.75 \\
&\int_M d\mu_M(x_1)\t^{\td l}\land \t^{\td n}\land \t^{\td m}\wh\Phi_i(x_1)\wh\Phi{}^\ast_j(x_1)=\gd M,\td l\td n\td m,ij, & \hbox{3-forms} \lb 5.76
\e
It is easy to see that in the case of a bosonic part of GSW model the situation is simpler because $M=S^2$ and all 3-forms are equal to zero.

For the remaining 1-forms and 2-forms one gets
\bg5.77
\int_{S^2}\sin\psi \,d\psi\,d\vf\, \t^\ta Y_{jm}(\psi,\vf)Y^*_{j'm'}(\psi,\vf)=\gd K,\td a,jmj'm',, \quad \ta=5,6,\\
\int_{S^2}\sin\psi \,d\psi\,d\vf\, \t^\ta\land \t^\tb Y_{jm}(\psi,\vf)Y^*_{j'm'}(\psi,\vf)=\gd L,\td a\td b,jmj'm',, \quad \ta,\tb=5,6,\ \ta\ne\tb \lb5.78 \\
\gd K,6,jmj'm', = d\vf\,\d_{jj'}\d_{mm'} = \t^6\d_{jj'}\d_{mm'}\,, \lb5.79 \\
\gd K,5,jmj'm',= d\psi\,\frac{\d_{mm'}}{4\pi}\sqrt{\frac{(2j+1)(2j'+1)(j-m)!\,(j'-m)!}{(j+m)!\,(j'+m)!}}\cdot \int_{-1}^1dx\,P^m_j(x)P^m_{j'}(x)\kern30pt \nn\\
\kern30pt {}=\frac{\t^5}{\sin\psi}\cdot\frac{\d_{mm'}}{4\pi}\sqrt{\frac{(2j+1)(2j'+1)(j-m)!\,(j'-m)!}{(j+m)!\,(j'+m)!}}
\cdot\int_{-1}^1 dx\,P^m_j(x)P_{j'}^m(x) \lb5.80 \\
\gd L,56,jmj'm', = \t^5\land \t^6 \frac{\d_{mm'}}{4\pi\sin\psi}\sqrt{\frac{(2j+1)(2j'+1)(j-m)!\,(j'-m)!}{(j+m)!\,(j'+m)!}}
\cdot\int_{-1}^1 dx\,P^m_j(x)P_{j'}^m(x). \lb5.80a
\e

The important problem in future calculations is to treat a part of a \ci\ \dv\ defined on~$M$ (or~$S^2$), i.e.
\beq5.81
\wt d(\Psi) + \wt\o{}^{\ta\tb} \wt\si_{\ta\tb}\land \Psi
\e
or
\beq5.82
\wt d(\ov\Psi)-\ov\Psi\land \wt\o^{\ta\tb}\wt\si_{\ta\tb}.
\e
One gets
\bg5.83
\wt d\Psi = \wt d(\Psi_\mu\T^\mu+\Psi_{\td m}\t^{\td m})= \xi_\ta(\Psi_\mu\T^\mu+\Psi_{\td m}\t^{\td m})\land \t^\ta \\
\wt d\ov\Psi = \wt d(\ov\Psi_\mu\T^\mu+\ov\Psi_{\td m}\t^{\td m})=\xi_\ta(\ov\Psi_\mu\T^\mu+\ov\Psi_{\td m}\t^{\td m})\land \t^\ta. \lb5.84
\e

The terms \er{5.72} in the case of a bosonic part of GSW model (20-\di al model) are easily obtained by taking $\ta,\tb=5,6$ and $m,l,n$ as G2 Lie
algebra indexes.

Further development will be done elsewhere. The important conclusion without further calculations is that a Velo--Zwanziger paradox is absent. This is
due to the \ia\ term in Eq.~\er{5.71} (the first case) and due to the term \er{5.72} (for the second and the third case). These terms influence \e s
of Rarita--Schwinger fields in such a way that first differential consequences are not algebraic constraints. They are differential \e s.

\section*{Appendix A}
\def\theequation{A.\arabic{equation}}
\setcounter{equation}0
In this Appendix we deal with \Ca\ (see Refs \cite{ax1}, \cite{ax2}) $C(1,n+3)$ or ${C(1,n+N_1+3)}$. Owing to de\cm\ rules for $C(1,n+3)$ ($C(1,n+N_1+3)$)
we write down a useful \rp ation for $\G^A$ (or~$\G^\tA$) in terms of~$\g^\mu$. It is well known that any \Ca s (see Refs \cite{ax1},~\cite{ax2}) can be
decomposed into a tensor products of the three \el ary \Ca s
\beq AA.1
\bga
C(0,1)=C \hbox{ --- complex numbers,}\\
C(1,0) = R \oplus R \\
C(0,2)= Q \hbox{ --- quaternions.}
\ega
\e
We have
\beq AA.2
C(1,n+3) = C(0,2) \ot C(1,n+1).
\e
We define \Ca\ $C(1,3)$ and we easily get
\beq AA.3
C(1,n+3) = \Bigl(\bigotimes_{i=1}^{[n/2]} C(0,2)\Bigr) \ot C(1,3) = \Bigl(\bigotimes_{i=1}^{[n/2]} H\Bigr) \ot C(1,3)
\e
or
\beq AA.4
C(1,n+n_1+3) = \Bigl(\bigotimes_{i=1}^{[(n+n_1)/2]} C(0,2)\Bigr) \ot C(1,3) = \Bigl(\bigotimes_{i=1}^{[(n+n_1)/2]} H\Bigr) \ot C(1,3).
\e
It is very well known that either
\bmlg
C(1,n+3)=C(1,n+4) \q (C(1,n+n_1+3)=C(1,n+n_1+4))\\
\hbox{iff } n+3=2l \q (n+n_1+3=2l)
\e
or
\bmlg
C(1,n+2)=C(1,n+3) \q (C(1,n+n_1+2)=C(1,n+n_1+3))\\
\hbox{iff } n+3=2l+1 \q (n+n_1+3=2l+1),
\e
$l \in N_1^\iy$.

Let $\g^\mu \in L(\C^4)$, $\mu=1,2,3,4$, be Dirac matrices obeying conventional relations
\beq AA.6
\bga
\{\g^\mu,\g^\nu\}=2\eta^\m\\
\eta^\m =\diag(-1,-1,-1,-1)\\
\g^5=\g^1\g^2\g^3\g^4, \q \g_5^2=1
\ega
\e
and let $\si_i\in L(\C^2)$, $i=1,2,3$, be Pauli's matrices obeying conventional relations as well
\beq AA.7
\bga
\{\si_i,\si_j\} = 2\d_{ij}\\
[\si_i,\si_j] = \ve_{ijk}\si_k.
\ega
\e

Let $I$ be $2\tm2$ unit matrix and let $J\in L(\C^4)$ be $4\tm4$ unit matrix. Thus we can perform a de\cm\
\beq AA.10
\G^\mu = \g^\mu \ot \Bigl(\bigotimes_{i=1}^{[n/2]} \si_1\Bigr)
\e
or
\beq AA.11
\G^\mu = \left(\begin{matrix} 0 &\ &\ldots &\ &\g^\mu\\
&&\udots&&\\ \g^\mu&&\ldots&&0 \end{matrix}\right).
\e
For $A\ne \mu$ ($\wt A\ne\mu$) one gets (for $n=2l$ or $n+n_1=2l$)
\beq AA.12
\bal
\G^{2p+1} &= iJ\ot \Bigl(\bigotimes_{i=1}^{p-2} I\Bigr)\ot\si_3\ot \Bigl(\bigotimes_{i=1}^{l-p+1}\si_1\Bigr)\\
\G^{2p+2} &= iJ\ot \Bigl(\bigotimes_{i=1}^{p-2} I\Bigr)\ot\si_3\ot \Bigl(\bigotimes_{i=1}^{l-p+1}\si_1\Bigr)
\eal
\e
where $4<2p+1<2p+2\le n+4=2l+2$ (or $4<2p+1<2p+2\le n+n_1+4=2l+2$). In the case $n=2l$ we define also a matrix
\bea AA.13
\G^{n+5} = i i^{3(l+1)} \prod_{A=1}^{n+4} \G^A = \g^5\ot \bigotimes_{i=1}^l \si_1= \G^{2l+5}\\
\G^{n_1+n+5} = i i^{3(l+1)} \prod_{A=1}^{n+n_1+4} \G^\tA = \g^5\ot \bigotimes_{i=1}^l \si_1= \G^{2l+5} \lb{AA.14}
\e
or
\beq AA.15
\G^{n+5}=\left(\begin{matrix} 0 &\ &\ldots &\ &\g^5\\
&&\udots&&\\ \g^5&&\ldots&&0 \end{matrix}\right)
\e
where $n=2l$, $l\in N_1^\iy$,
\beq AA.16
\G^{n+n_1+5}=\left(\begin{matrix} 0 &\ &\ldots &\ &\g^5\\
&&\udots&&\\ \g^5&&\ldots&&0 \end{matrix}\right),
\e
$n_1+n=2l$, $l\in N_1^\iy$.

If $n=2l+1$ we have
\beq AA.17
\ov\G{}^A= \G^A, \q A=1,2,\dots,2l+4, \q \ov\G{}^{n+4}=\G^{2l+5}=\left(\begin{matrix} 0 &\ &\ldots &\ &\g^5\\
&&\udots&&\\ \g^5&&\ldots&&0 \end{matrix}\right).
\e
Similarly, if $n+n_1=2l+1$ we have
\beq AA.18
\ov\G{}^\tA= \G^\tA, \q \tA=1,2,\dots,2l+4, \q \ov\G{}^{n+n_1+4}=\G^{2l+5}=\left(\begin{matrix} 0 &\ &\ldots &\ &\g^5\\
&&\udots&&\\ \g^5&&\ldots&&0 \end{matrix}\right).
\e
It is easy to check that
\bg AA.19
(\G^{2l+5})^2= -1, \q \{\wt\G{}^A,\G^{2l+5}\}=0 \qh{for} A\ne 2l+5,\\
B = \ov B\ot \Bigl(\bigotimes_{i=1}^{[n/2]}\si_1\Bigr), \q \g^{\mu+}=\ov B\g^\mu \ov B{}^{-1}. \lb{AA.20}
\e
The same we get for $n+n_1+4$ case
\beq AA.21
\bga
\{\wt\G{}^\tA,\G^{2l+5}\}=0 \qh{for} \tA\ne 2l+5,\\
B = \ov B\ot \Bigl(\bigotimes_{i=1}^{[(n+n_1)/2]}\si_1\Bigr).
\ega
\e
Usually $B=\g^4$.
\bg AA.22
\ov\G^{21}=\left(\begin{matrix} 0 &\ &\ldots &\ &\g^5\\
&&\udots&&\\ \g^5&&\ldots&&0 \end{matrix}\right)\\
\bigl(\ov\G{}^{21}\bigr)^2 = 1 \lb{AA.22a}
\e

We can proceed in a little different way:

Generalized Dirac matrices are defined by the relations
\beq AA.23
\{\G^A,\G^B\}=2\eta^{AB} \qh{or} \{\G^{\tA},\G^{\tB}\}=2\eta^{\tA\tB}
\e
where
\bea AA.24
\eta^{AB}&=\mathop{\rm diag}\{-1,-1,-1,1,\underbrace{-1,\dots,-1}_n\}\\
\eta^{\tA\tB}&=\mathop{\rm diag}\{-1,-1,-1,1,\underbrace{-1,\dots,-1}_{n_1},\underbrace{-1,\dots,-1}_n\}. \lb{AA.25}
\e
For $(n+4)$ or $(n+n_1+4)$ equal to $2l+2$ (the even case) we define
\beq AA.26
\bal \G^{4\pm}&=\tfrac12(\pm\G^4+\G^1),\\
\G^{\bar A\pm}&=\tfrac12(\G^{2\bar A}\pm i\G^{2\bar A+1}), \q \ov A=1,\dots,l.
\eal
\e
It is easy to show
\beq AA.27
\bga
\{\G^{\bar A\pm},\G^{\bar B\pm}\}=-\d^{\bar A\bar B}\\
\{\G^{\bar A+},\G^{\bar B+}\}=\{\G^{\bar A-},\G^{\bar B-}\}=0.
\ega
\e
In particular
\beq AA.28
(\G^{\bar A+})^2=(\G^{\bar A-})^2=0.
\e
In this way we always have a spinor $\Ps_0$ \st
\beq AA.29
\G^{\bar A-}\Ps_0=0
\e
for all $\ov A$. We get all possible spinors acting on $\Ps_0$ by $\G^{\bar A+}$.
We get $2^{l+1}$ such spinors (a~full \rp ation). $\G^A$ or $\G^\tA$ can be
derived in such a base by using iterative method.

In the case of $2l+3$ (an odd case) we should have
\beq AA.30
\G^{2l+3}=i^{-(l+1)}\G^1 \cdots \G^{2l+2}
\e
\st
\beq AA.31
(\G^{2l+3})^2=-1, \q \{\G^{2l+3},\G^{\bar{\bar A}}\}=0, \q \ov{\ov A}=1,\dots,2l+2.
\e
It is easy to define a basis of spinors for both cases. Let $\z=(\z_1,\dots,
\z_l)$, $\z_{\bar A}=\pm\frac12$,
\beq AA.32
\Ps_\z = \Bigl(\prod_{\bar A=0}^l (\G^{(l+\bar A)})^{\z_{(l+\bar A)}+1/2} \Bigr)\Ps_0.
\e
$\G^{2l+3}$ in the even case distinguishes between two classes of spinors
\beq AA.33
\bal \G^{2l+3}\Ps_\z &= +\Ps_\z \qh{($2^l$-\di---first \rp ation)}\\
\G^{2l+3}\Ps_\z &= -\Ps_\z \qh{($2^l$-\di---second \rp ation)}
\eal
\e
In the odd case we have only one \rp ation of $2^{l+1}$-\di.

We can introduce also generators of $\SO(1,3+n)$ or $\SO(1,3+n_1+n)$ algebra
\beq AA.34
\bga
\ov\si{}^{AB} \qh{or} \ov\si{}^{\tA\tB}\\
\ov\si{}^{AB}=\frac 14[\G^A,\G^B]\\
\ov\si{}^{\tA\tB}=\frac 14[\G^\tA,\G^\tB].
\ega
\e
We have of course
\beq AA.35
\bga
{}[\ov\si{}^{MN},\ov\si{}^{RS}] = - [\eta^{NS}\ov\si{}^{MR}+ \eta^{RN}\ov\si{}^{SM}
+\eta^{MR}\ov\si{}^{NS}+\eta^{SM}\ov\si{}^{RS}]\\
[\ov\si{}^{\td M\td N},\ov\si{}^{\td R\td S}] = - [\eta^{\td N\td S}\ov\si{}^{\td M\td R}+ \eta^{\td R\td N}\ov\si{}^{\td S\td M}
+\eta^{\td M\td R}\ov\si{}^{\td N\td S}+\eta^{\td S\td M}\ov\si{}^{\td R\td S}].
\ega
\e
Our spinors transform as
\beq AA.36
\bal
\PS&\to \exp\bigl(\tfrac12 \ve_{AB}\wh\si{}^{AB})\Ps \\
\hbox{or}\q \PS&\to \exp\bigl(\tfrac12 \ve_{\tA\tB}\ov\si{}^{\tA\tB})\Ps \\
&\ve_{AB}=-\ve_{BA}\\
&\ve_{\tA\tB}=-\ve_{\tB\tA}.
\eal
\e
We also have
\beq AA.37
\bal
(\ov\si{}^{AB})^+ \G^4&=\G^4 \ov\si{}^{AB}\\
(\ov\si{}^{\tA\tB})^+ \G^4&=\G^4 \ov\si{}^{\tA\tB}.
\eal
\e

In our particular cases with or without \sn\ \s y breaking we get our
matrices using ordinary Dirac matrices and their tensor products with some
special matrices. One gets for \ci\ \dv s
\beq AA.38
\bal
\wt D\Ps&=d\Ps + \tfrac12\wt\o_{AB}\ov \si{}^{AB}\Ps\\
\wt D\Ps&=d\Ps + \tfrac12\wt\o_{\tA\tB}\ov \si{}^{\tA\tB}\Ps.
\eal
\e
Moreover, we use as before (see Ref.~\cite8)
\beq AA.39
\bal
\gv{\wt D}\Ps &= \hor \wt D\Ps = \gv{d}\Ps+\tfrac12\hor(\wt\o_{AB})\ov\si{}^{AB}\Ps\\
\gv{\wt D}\Ps &= \hor \wt D\Ps = \gv{d}\Ps+\tfrac12\hor(\wt\o_{\tA\tB})\ov\si{}^{\tA\tB}\Ps
\eal
\e
and also
\beq AA.40
\bal
\wt D\ov\Ps &= d\ov\Ps - \tfrac12\wt\o_{AB}\ov\Ps \ov\si{}^{AB}\\
\hbox{or} \q \wt D\ov\Ps &= d\ov\Ps
- \tfrac12\wt\o_{\tA\tB}\ov\Ps \ov\si{}^{\tA\tB}
\eal
\e
where
\beq AA.41
\ov\Ps = \Ps^+ \G^4
\e
and similarly
\beq AA.42
\bal
\gv{\wt D}\ov\Ps &= \gv{d}\ov\Ps - \tfrac12\hor(\wt\o_{AB})\ov\Ps
\ov\si{}^{AB}\\
\hbox{or} \q \gv{\wt D}\ov\Ps &= \gv{d}\ov\Ps
- \tfrac12\hor(\wt\o_{\tA\tB})\ov\Ps \ov\si{}^{\tA\tB}.
\eal
\e
$\wt\o_{AB}$ and $\wt\o_{\tA\tB}$ are \LC \cn s defined on~$P$ \wrt a \s ic
part of metrics $\g_\(AB)$ and $\g_\(\tA\tB)$.

How does an iterative method for a construction of $\G$ matrices work? Let us
suppose we have ordinary Dirac matrices $\g^\mu$ and let us define
\beq AA.43
\bal
\G^\mu &= \g^\mu \otimes \left(\begin{matrix} -1 &\ & 0\\ 0 && 1
\end{matrix}\right), \q \mu=1,2,3,4, \\
\G^5 &= J \otimes \left(\begin{matrix} 0 &\ & 1 \\ 1 && 0
\end{matrix}\right), \\
\G^6 &= J \otimes \left(\begin{matrix} 0 &\ & -i \\ i && 0
\end{matrix}\right) , \q J \hbox{ an identity matrix, }4\tm4.
\eal
\e
Next step
\beq AA.44
\bal
\ov\G{}^A &= \G^A \otimes \left(\begin{matrix} -1 &\ & 0 \\ 0 && 1
\end{matrix}\right) , \q A=1,2,3,4,5,6,\\
\ov\G{}^7 &= I_6 \otimes I_6 \otimes \left(\begin{matrix} 0 &\ & 1\\ 1 && 0
\end{matrix}\right),\\
\ov\G{}^8 &= I_6 \otimes I_6 \otimes \left(\begin{matrix} 0 &\ & -i \\ i && 0
\end{matrix}\right), \q I_6 \hbox{ an identity matrix, }6\tm6.
\eal
\e

For a future convenience let us apply our formalism in 20-\di al case (i.e.\ for GSW model). One gets
\bg AA.45
\{\G^A,\G^B\} = 2\eta^{AB}\\
\eta^{AB}=\diag(-1,-1,-1,1,-1,-1,\underbrace{-1,-1,\dots,-1}_{14}) \lb{AA.46}
\e
Let us define
\bg AA.47
\G^A=\wt\G{}^A \ot I, \q 1\le A\le 4, \\
\G^A=\wt\G{}^5 \ot \G^{A-4}, \q 4 < A \le 20, \lb{AA.48}
\e
or
\bea AA.49
\G^\mu &= \g^\mu \ot I_8, \q 1\le\mu\le 4,\\
\G^5 &= \g^5, \lb{AA.50} \\
\G^{r+4}&= \g^5 \ot \G^r, \q 1<r\le 16. \lb{AA.51} \\
\ov\G{}^{21}&=-(-i)^9\cdot \G^1 \cdots \G^{20} =  i\cdot \G^1 \cdots \G^{20} \lb{AA.52} \\
\ov\G{}^{21}&=\g^5 \ot \ov\G{}^{17} \nn \\
\ov\G{}^{17}&=(-i)^8 \G^1 \cdots \G^{16} = \G^1 \cdots \G^{16},\lb{AA.53}
\e
$I_8$ --- identity matrix $8\tm8$, $\g^\mu$ --- $4\tm 4$ matrices as usual, $\G^A$ --- $10\tm 10$ matrices.

\section*{Appendix B}
\def\theequation{B.\arabic{equation}}
\setcounter{equation}0
In this appendix we give some formulas for \ci\ \dv\ used by us in the paper (see Refs \cite{ax4}, \cite{ax5}, \cite{ax6}, \cite{8}). We have
\bml AB.1
\cD\Psi = \gv{\ov{d\Psi}} + \hor(\gd \wt\o,\tA,\tB,)\gd\wh\si,\tB,\tA,\Psi
=\bigl(\gv d\Psi+\xi_\ta \Psi\t^\ta + \F^a_\ta X_a\Psi\t^\ta\bigr)
+\hor(\gd\wt\o,\a,\b,\gd\wh\si,\b,\a,)\Psi\\ {}+\hor(\gd\wt\o,\ta,\tb,\gd\wh\si,\tb,\ta,)\Psi
+\hor(\gd\wt\o,a,b,\gd\wh\si,b,a,)\Psi +\hor(\gd\wt\o,\a,b,)\gd\wh\si,b,\a,\Psi
+\hor(\gd\wt\o,a,\b,)\gd\wh\si,\b,a,\Psi +\hor(\gd\wt\o,a,\tb,)\gd\wh\si,\tb,a,)\Psi,
\e
$\gv{\ov{d\Psi}} = \hor(d\Psi)$, $\gv{\ov d}\ov\Psi=\hor(d\ov\Psi)$,
\bml AB.2
\cD\ov\Psi = \gv{\ov{d\Psi}} - \hor(\gd \wt\o,\tA,\tB,)\ov\Psi \gd\wh\si,\tB,\tA,
= \bigl(\gv d \Psi+\xi_\ta \ov\Psi \t^\ta - \F^a_\ta \ov\Psi X_a\t^\ta\bigr)
-\hor(\gd \wt\o,\a,\b,)\ov\Psi\gd\wh\si,\b,\a,\\
{}-\hor(\gd \wt\o,\ta,\tb,)\ov\Psi\gd\wh\si,\tb,\ta,
-\hor(\gd \wt\o,a,b,)\ov\Psi\gd\wh\si,b,a, -\hor(\gd \wt\o,\a,b,)\ov\Psi\gd\wh\si,b,\a,
-\hor(\gd \wt\o,\ta,b,)\ov\Psi\gd\wh\si,b,\ta, -\hor(\gd \wt\o,a,\tb,)\ov\Psi\gd\wh\si,\b,a,.
\e

\er{AB.1} and \er{AB.2} cover both cases with and without \ssb. For GSW model in our approach we have
\bml AB.3
\cD\Psi = \gv{\ov{d\Psi}} + \hor(\gd \wt\o,\tA,\tB,)\dg\wh\si,\tA,\tB,\Psi
=\bigl(\gv d\Psi + \pa_5\Psi\t^5 + \pa_6\Psi\t^6 + \F_5\Psi\t^5 + \F_6\Psi\t^6\bigr)\\
{}+\hor(\gd \wt\o,\a,\b,)\dg\wh\si,\a,\b, \Psi
+2\hor(\gd \wt\o,5,6,)\dg\wh\si,5,6, \Psi
+\hor(\gd \wt\o,a,\b,)\dg\wh\si,a,\b, \Psi
+\hor(\gd \wt\o,\a,b,)\dg\wh\si,\a,b, \Psi
+\hor(\gd \wt\o,a,5,)\dg\wh\si,a,5, \Psi\\
{}+\hor(\gd \wt\o,a,6,)\dg\wh\si,a,6, \Psi
+\hor(\gd \wt\o,5,b,)\dg\wh\si,5,b, \Psi
+\hor(\gd \wt\o,6,b,)\dg\wh\si,6,b, \Psi
+\hor(\gd \wt\o,a,b,)\dg\wh\si,a,b, \Psi
\e
\bml AB.4
\cD\ov\Psi = \gv{\ov{d}}\Psi - \hor(\gd \wt\o,\tA,\tB,)\ov\Psi\dg\wh\si,\tA,\tB,
=\bigl(\gv d\ov\Psi + \pa_5\ov\Psi\t^5 + \pa_6\ov\Psi\t^6 - \ov\Psi\F_5\t^5 - \ov\Psi\F_6\t^6\bigr)\\
{}-\hor(\gd \wt\o,\a,\b,)\ov\Psi\dg\wh\si,\a,\b,
-2\hor(\gd \wt\o,5,6,)\ov\Psi\dg\wh\si,5,6,
-\hor(\gd \wt\o,a,\b,)\ov\Psi\dg\wh\si,a,\b,
-\hor(\gd \wt\o,\a,b,)\ov\Psi\dg\wh\si,\a,b,
-\hor(\gd \wt\o,a,5,)\ov\Psi\dg\wh\si,a,5, \\
{}-\hor(\gd \wt\o,a,6,)\ov\Psi\dg\wh\si,a,6,
-\hor(\gd \wt\o,5,b,)\ov\Psi\dg\wh\si,5,b,
-\hor(\gd \wt\o,6,b,)\ov\Psi\dg\wh\si,6,b,
-\hor(\gd \wt\o,a,b,)\ov\Psi\dg\wh\si,a,b, .
\e

$\gd\wt\o,\tA,\tB,$ is a \LC \cn\ generated by a \s ic part if a tensor $\g_\(\tA\tB)$ on~$P$. $\gd\wt\o,\a,\b,$ is a \LC \cn\ generated by
$g_\(\a\b)$ on a \spt~$E$.  We have also a \ci\ \dv\ of spinors on a sphere~$S^2$. One gets
\bea AB.5
\wt\n_5 \Psi &= \wt\n_\psi \Psi = \pap{}\psi \Psi + \dgd\wt\G,6,5,\psi, \dg\wh\si,5,6,\Psi + \dgd\wt\G,5,6,\psi, \dg\wh\si,6,5,\Psi \\
\wt\n_6 \Psi &= \wt\n_\vf \Psi = \pap{}\vf \Psi + \dgd\wt\G,6,5,\vf, \dg\wh\si,5,6,\Psi + \dgd\wt\G,6,5,\vf, \dg\wh\si,6,5,\Psi \lb{AB.6} \\
\wt\n_5 \ov\Psi &= \wt\n_\psi \ov\Psi = \pap{}\psi \ov\Psi - \ov\Psi\dg\wh\si,5,6,\gd\wt\G,5,6\psi, - \ov\Psi\dg\wh\si,6,5, \dgd\wt\G,5,6,\psi, \lb{AB.7} \\
\wt\n_6 \ov\Psi &= \wt\n_\vf \ov\Psi = \pap{}\vf \ov\Psi - \ov\Psi\dg\wh\si,5,6,\gd\wt\G,5,6\vf,  - \ov\Psi\dg\wh\si,6,5,\gd\wt\G,6,5\vf,  \lb{AB.8}
\e
$\gd\wt\G,6,5\vf,$ and so on are Christoffel symbols for a \cn\ on~$S^2$. One gets
\beq AB.9
\bga
\gd\wt\G,5,66, -\sin\psi \cos\psi\\
\gd\wt\G,6,56,=\gd\wt\G,6,65,=\cot \psi.
\ega
\e
The remaining Christoffel symbols are zero. A metric tensor on~$S^2$ is defined
\beq AB.10
g_\(\ta\tb) = r^2\left(\begin{matrix} -1 &\q& 0 \\ 0 && -\sin^2\psi \end{matrix} \right)
\e
and the inverse of $g_\(\ta\tb)$:
$$
g^{\ta\tb} = \frac1{r^2}\left(\begin{matrix} -1 &\q& 0 \\ 0 && -\frac1{\sin^2\psi} \end{matrix} \right).
$$

\section*{Appendix C}
\def\theequation{C.\arabic{equation}}
\setcounter{equation}0
In this appendix we give some \el s of the Atiyah--Singer index theorem (see Refs \cite{aw1}, \cite{aw2}). The Atiyah--Singer index theorem
gives an equality between two types of indexes of an elliptic operator defined on a compact manifold~$X$. The first is known as an analytical
index and the second as a topological index. An analytical index for an elliptic operator~$D$ is defined as follows
\beq AC.1
{\rm Index}(D)=\dim {\rm Ker}(D) - \dim {\rm Coker}(D)=\dim {\rm Ker}(D)-\dim {\rm Ker}(D^+)
\e
where $D^+$ is an adjoint operator for $D$.

${\rm Ker}(D)$ is defined as the space of \so s
\beq AC.2
Df=0.
\e
$D$ is of course a differential operator between two smooth vector bundles $E,F$ on a compact manifold~$X$,
\beq AC.3
D:E\to F.
\e

A topological index of an elliptic differential operator is given by
\beq AC.4
{\rm Topological\ index}(D)=(-1)^n \langle{\rm ch}({\rm s}(D)) \cdot {\rm Td}(T_C X),[X]\rangle
\e
where $n$ is the \di\ of the manifold~$X$, s$(D)$ is the symbol of the operator~$D$, ch is the Chern character,
$T_CX$ is the complexified tangent bundle of~$X$, ``$\cdot$'' is the cup product, $[X]$~is a \fn\ class of~$X$ and
\beq AC.5
\langle\o,[X]\rangle=\int_X \o.
\e
If the operator $D$ is given by the formula
\beq AC.6
D=\sum_{|\a|\le m} a_\a(x)D^\a, \q
D^\a=\pap{^{\a_1}}{x_1^{\a_1}} \cdot \pap{^{\a_2}}{x_2^{\a_2}} \cdots \pap{^{\a_n}}{x_n^{\a_n}} , \q \a=(\a_1,\dots,\a_n),
\e
s$(D)$ is given by
\beq AC.7
{\rm s}(D)(x,p) = \sum_{|\a|=m}a_\a(x)p^\a.
\e
In this way \er{AC.4} is defined in pure topological terms. The index theorem states that both indexes are equal. We also have
\beq AC.5a
{\rm Index}(D)=\Tr(e^{-tD^+D}) - \Tr(e^{-tDD^+}), \q t\in\R.
\e
The differential operator $D$ has a pseudoinverse which is a Fredholm operator.

\advance\abovedisplayskip by-2pt \advance\belowdisplayskip by-2pt
In our case $D$ is $\Dint$ defined on a compact group manifold. It means it is a Dirac operator. Due to the index theorem an analytical index
is equal to a topological \iv t. Thus any smooth deformation of~$D$ cannot change the value of the index. This means that any type of smooth \LC (or
even metric) \cn\ on~$X$ is not able to change the value of an index if we change
\beq AC.6a
\pa_a \to \wt\n_a.
\e
This is valid also if we introduce a nontrivial ``gauge'' \dv\
\beq AC.7a
\pa_a \to \gv{\n_a}
\e
or even both \dv s at once
\beq AC.8
\pa_a \to \gv{\wt\n_a}.
\e
This allows us to pose a problem of chiral \fe\ in our theory on a stable mathematical ground which we do in the paper (see Refs \cite{ay1},
\cite{ay2}, \cite{ay3}, \cite{ay4}, \cite{ay5}, \cite{ay6}).

For in our theory we have to do with group manifolds ($G$, $H$ or $G2$) we can consider $\wh X_a$ (differential operators on the manifold,
i.e.\ left \iv t vector fields) in the place of $\pa_a$. In this way we have to do with anholonomic frame (a~group in general is \nA)
and in the place of $\pa_a$ we have $\wh X_a$. The \ci\ \dv s $\wt\n_a$, $\gv{\n_a}$ or $\gv{\wt\n_a}$ are defined in this anholonomic frame.
Moreover, the zero modes condition should be redefined in the \fw\ way
\beq AC.12
\G^a \wh X_a \Psi=0.
\e
Thus we have to do with a smooth deformation of condition \er{5.4}. This is true also in the case of \ci\ \dv\ written in an
anholonomic frame.

Due to the index theorem this does not change a value of a topological index because it results in a smooth deformation of a differential operator.
In particular, the space of $\F_i$ or $\wh\F_i$ functions will be the same. Moreover, we have a gauge condition \er{AC.12} for our spinor fields.

\section*{Conclusions}
In the paper we consider a problem of a geometrical coupling of \fe s to the bosonic \un\ of \fn\ \ia s including gravity. We get Yukawa term
in the case of \ssb\ and \Hm. The further research consists in placing of existing \fe\ families in the scheme, possibly getting some predictions
of new kind of \fe s and their masses and mixing angles. In the case of the \NK\ we consider also a different approach (see Refs \cite{r1}, \cite{r2},
\cite{r3}). However, now we consider the present one as more interesting.

Let us give the following comment. The standard \KK Theory uses a \LC \cn\ on a naturally metrized gauge principale fiber bundle
(see Ref.~\cite{12}). The \NK\ uses non-Riemannian affine \cn\ on a non\s ically metrized gauge principal fiber bundle
(see Ref.~\cite{1}). Both geometries are \ti{different\/}. Moreover, there is a~\cn\ between these two theories via Bohr correspondence \pp\
because an affine \cn\ of the \NK\ contains a \LC part. In this way if the skew\s ic part of the metric is zero, it collapses to standard
\KK Theory (roughly speaking). The \fe\ fields in the standard \KK Theory (as in our previous papers) are coupled via new types of gauge \dv s which
are horizontal parts of exterior \ci\ \dv s of \LC \cn\ on a metrized gauge principal fiber bundle. We do not consider here (i.e., in standard
\KK Theory) zero modes of Dirac operator for \fe\ fields on a group manifold. In the case of the \NK\ we couple zero modes of the \fe\ fields via
the same new types of gauge \dv s as in standard \KK Theory. Moreover, in the case of the \NK\ with \ssb\ and \Hm\ we are getting completely new
results in comparison to the standard one. This is due to the more complicated \sc e of \ssb\ and Higgs' sector going to more extended Yukawa
coupling for \fe s. In the simplest case considered here by us (i.e.\ bosonic part of GSW model in the \NK) we get Yukawa coupling for \fe s and
a~theory with masses of $W^\pm,Z^0$ and Higgs' boson agreed with an experiment. This is impossible in the standard (\s ic) \KK Theory.

We think that this comment could give a clear statement what distinguishes the \NK\ from standard \KK Theory, and in what respect the \fe\ part
is different.

In 1977 we considered an idea to use a torsion of a \cn\ defined on a fiber bundle manifold (6-\di al, or $(n+5)$-\di al, where $n$ is a \di\ of a gauge group).
The bundle has been defined over a metrized \elm c bundle (a~bundle over a~bundle). The metrization has been achieved according to the
Trautman--Tulczyjew idea (see Ref.~\cite{12}). We wanted to geometrize Higgs' field, \ssb\ and \Hm.

This idea has been developed further in order to get self-\ia\ terms in a scalar \cvt\ derived for a metric \cn\ (moreover, with non-vanishing
torsion) for a Higgs' field. The Higgs' field has been interpreted as~$A_5$ (the fifth \cd\ of an \elm c \pt). If we suppose that $A_5=Q$ (a~scalar
field) generates a torsion of a \cn\ on a metrized fiber bundle we get a self-interacting \pt\ of the scalar field~$Q$. The torsion should have
a~special dependence on the scalar field~$Q$ (a~Higgs' field). We considered also $(n+5)$-\di al case with a multiplet of scalar fields~$Q^a$,
where $Q^a=\gd A,a,5,$ is a part of a gauge field over additional \di s. We abandon this idea as useless and do not develop it further. Moreover,
an idea of $A_5$,~$\gd A,a,5,$ as Higgs' fields is correct and it was developed in our future papers.

This is a historical remark.

\section*{Acknowledgement}
I would like to thank Professor B. Lesyng for the opportunity to carry out
computations using Mathematica\TM~9\footnote{Mathematica\TM\ is the
registered mark of Wolfram Co.} in the Centre of Excellence
BioExploratorium, Faculty of Physics, University of Warsaw, Poland.

\end{document}